\documentclass{article}

\usepackage{geometry}
\usepackage{graphicx}  
\usepackage{amsmath}   
\usepackage[compress]{cite}
\usepackage{amssymb}   
\usepackage{bm} 
\usepackage{dcolumn}
\usepackage{color}
\usepackage{mathrsfs}
\usepackage{amsfonts}
\usepackage{varioref}
\usepackage{csquotes}
\usepackage{hyphenat}
\usepackage[caption=false]{subfig}
\RequirePackage{flushend}
\usepackage[table]{xcolor}
\usepackage{array,siunitx}
\usepackage{changepage}
\newcolumntype{M}{>{\centering\arraybackslash}m{2cm}}

\usepackage{array}
\usepackage{makecell}

\RequirePackage[numbers,sort&compress]{natbib}

\RequirePackage[colorlinks,citecolor=blue,urlcolor=magenta,linkcolor=blue]{hyperref}

\def\P{Mathisson-Pirani}
\def\T{Tulczyjew-Dixon}
\def \NW {Newton-Wigner}
\def \MP {Mathisson-Papapetrou}
\def \SSC {spin supplementary condition}
\def \ISCO {innermost stable circular orbit}
\labelformat{section}{Section #1} 
\labelformat{subsection}{Section #1} 
\labelformat{subsubsection}{Section #1}
\labelformat{subsubsubsection}{Section #1}
\labelformat{equation}{Eq.~(#1)} 
\labelformat{figure}{Fig.~(#1)} 
\labelformat{subfigure}{Fig.~(\thefigure#1)} 
\labelformat{table}{Tab.~(#1)} 
\labelformat{appendix}{Appendix #1}

\title{Off-equatorial stable circular orbits for spinning particles }
\author{Sajal Mukherjee\footnote{sm13ip029@iiserkol.ac.in}~$^{1}$ and K. Rajesh Nayak\footnote{rajesh@iiserkol.ac.in}~$^{1,2}$\\
$^{1}${\small{Department of Physical Sciences, IISER-Kolkata, Mohanpur-741246, India.}}\\
$^{2}${\small{Center of Excellence in Space Sciences India, IISER-Kolkata, Mohanpur-741246, India.}}}

\begin{document}
\maketitle

\begin{abstract}

In this article, we investigate the motion of a spinning particle at a constant inclination, different from the equatorial plane, around a Kerr black hole. We mainly explore the possibilities of stable circular orbits for different spin supplementary conditions. The Mathisson-Papapetrou's equations are extensively applied and solved within the framework of linear spin approximation. We explicitly show that for a given spin vector of the form $S^{a} = \left(0,S^r,S^{\theta},0\right)$ , there exists an unique circular orbit at $(r_c,\theta_c)$ defined by the simultaneous minima of energy, angular momentum and Carter constant. This corresponds to the Innermost Stable Circular Orbit (ISCO) which is located on a non-equatorial plane. We further establish that the location ($r_c,\theta_c$) of the ISCO for a given spinning particle depends on the radial component of the spin vector ($S^r$) as well as the angular momentum of the black hole ($J$). The implications of using different \SSC s are investigated.

\end{abstract}

\section{Introduction}
In this era of gravitational wave astronomy, modeling the relativistic two body problem is of vital interests. Due to the non-linearity of the system, the relativistic two body problem can not be solved exactly within the framework of Einstein's equations and one has to rely on approximations or numerical solutions. Two of well known approximations are : Post Newtonian or PN approximation \cite{Blanchet:2013haa} and Effective one body or EOB approach \cite{Buonanno:1998gg,Damour:2008yg}. Even if these methods are good enough to explain the system in the linear region where the relativistic effects are small, in the non-linear domain these approximations breakdown. This is when the numerical tools become inevitable. In recent years, Numerical relativity has become an essential part of gravitational wave astronomy \cite{Pretorius:2005gq,Baker:2005vv}. With the advance of computational techniques, the numerical methods are improving in a rapid rate and new physics emerges every moment beyond analytic understandings \cite{Garfinkle:2016lcu}. Besides these successes, Numerical relativity has its fair share of limitations. In particular, Numerical relativity is not vary efficient when the mass ratios become extremely large or small in a binary system  \cite{Cardoso:2014uka}. In these scenarios, the approximate techniques such as the effective one body formalism are useful.

With these motivations, we study the motion of spinning objects within the pole-dipole approximation in curved spacetime. In actual astrophysical situations, this corresponds to the orbiting of a compact object, representing a black hole or neutron star, of mass $M_1$ ($\sim $ few solar mass) around a massive black hole of mass $M_2$ (  $\sim 10^4 M_{\odot}$ to $10^6 M_{\odot}$) such that $M_2 >>M_1$. This is usually referred as extreme mass ratio inspiral (EMIR) which is a promising source of gravitational waves for proposed space based detectors such as LISA \cite{Sopuerta:2012hg}. In these scenarios, the internal structures of the orbiting body is approximated to a dipole and all other higher moments are ignored. Even in this lowest order approximation, pole-dipole particles can have striking deviation from a geodesic trajectory. In curved spacetime, the orbits of these particles are described by the \MP~equations \cite{Mathisson:1937zz,Papapetrou:1951pa}. These equations has a long and substantial history spanning over few decades, an extensive literature survey can be found in \cite{Semerak:1999qc}. We also refer our readers \cite{papapetrou1955pol,Hojman:1976kn,Prasanna:1989ry,Suzuki:1996gm,Plyatsko:2013xza} for further insights. Though the exact solutions of these equations are extremely complicated, there are several approach for solving them with suitable approximations. In the present article, we have used the linear spin approximations in which we write the \MP~equations upto the linear order in spin and neglect higher order terms. With this approximation, one can solve the orbit equations for an arbitrary inclination angle ($\theta$) and investigate the possible existence of circular orbits on  $\theta=\textit{constant}$ planes. In passing, we note that for a single pole particle, there is no circular orbit, either stable or unstable, can exist in the Kerr spacetime at constant inclination angle except for $\theta=\pi/2$ \cite{de1979non}.

In general, for binaries with unequal masses, especially with extreme mass ratios, the system undergoes a precession and orbits start to wobble along the off-equatorial planes whenever the angular momentums are not aligned with each other. Here, we have shown that under specific conditions, it is possible to have orbits without any wobbling. Apart from the wobbling in the off-equatorial directions, relativistic orbits can precess while confined to a particular orbital plane. This is usually known as the Periastron precession and in our solar system, it is called Perihelion precession \cite{einstein1915erklarung,wald2010general}. As in the present context we are only concentrating on the circular orbits, Periastron precession would identically vanish. What we consider, are families of stable circular orbits with an ISCO. We investigate the properties associated with these orbits for different \SSC s. The occurrence of such conditions are natural as the motion of a spinning object depends on the choice of a reference point and each choice would lead to a distinct \SSC. In this article, we mainly concentrate on \P~\cite{Pirani:1956tn} or \T~\SSC~\cite{tulczyjew1959motion} and \NW~\SSC~\cite{Newton:1949cq}. Even if, \T~and \P~conditions are distinct from each other for the exact \MP~equations, they both merge in the limit of linear spin approximation.

The rest of the manuscript is organized as follows. In \ref{sec:EOM}, we elaborately describe the motion of a spinning particle for different spin supplementary conditions while exclusively using the linear spin approximation. We then introduce the conserved quantities for a spinning particle such as, energy, angular momentum and Carter constant. \ref{sec:R_Theta} is devoted to study the motion of spinning particles numerically and discuss the existence of circular orbits at constant altitude. In \ref{sec:Stability}, we discuss the stability of these circular orbits located in the non-equatorial planes and further carry out detailed numerical analysis to investigate any possible existence of ISCO for various rotation parameters of the black hole. Finally, we close the article with a brief remark in \ref{sec:Discussion}. 

\textit{Notation and Conventions :} Throughout the paper we have used the $(-,+,+,+)$ signature with the fundamental constants $c=1=G$. In addition, any four vector $X^{\mu}$ projected on the tetrad frame is given as $X^{(\mu)}=e^{(\mu)}_{~\nu}X^{\nu}$, where $e^{(\mu)}_{~\nu}$ is the tetrad field.

\section{Basic equations for a Spinning Particle}\label{sec:EOM}
The trajectory of a single-pole test particle in a gravitational field is given by the geodesic equation which is obtained by setting the acceleration to zero. Unlike Newtonian gravity, general relativity does not treat gravity as a force, instead, depicts it as an inbuilt manifestation of the spacetime itself. This is, in fact, one of the very basic postulate of Einstein's gravity \cite{meaning}. Motion of a particle can deviate from geodesic trajectories in presence of a force. This force can be external or  internal,  if the particle has higher order mass multipoles. In a realistic situation, the astrophysical objects are expected to have complex internal structure. The first order correction to the single-pole test particle would be to add a dipole moment along with the monopole  to incorporate the internal angular momentum of the object. By dipole moment, we mean the center of mass of the spinning body in its rest frame does not coincide with the observed center of mass in the observer's frame. This is because, for a spinning particle in curved spacetime, in general, the center of mass is observer dependent \cite{Steinhoff:2015ksa}. The motion of these particles are described by the \MP~equations and for a four momentum, $P^a$ and spin tensor, $S^{a b}$, these can be written as:
\begin{equation}
\dfrac{DP^a}{d\tau} = -\dfrac{1}{2}R^a_{~b c d}U^b S^{c d}, \qquad {\rm and } \qquad \dfrac{DS^{a b}}{d\tau} = P^a U^b-P^b U^a.
\end{equation}
Here, $U^a$ is the four velocity of the particle and $R^{a}_{~b c d}$ is the Riemann curvature tensor. For a limiting case of $S^{ab}\rightarrow 0$, one gets back the geodesic equations, $i.e.$, acceleration, $a^{i}=U^{b}\nabla_b U^i=0$. The coupling of the spin tensor with the background geometry contribute to an acceleration and hence, the particle deviates from the usual geodesic trajectory.

In the case of spinning particles, the four momentum and four velocity are not proportional to each other. This will lead to total 14 unknown variables (four for each velocity and momentum and six for antisymmetric spin tensor), while we have only 10 equations in hand. In order to solve this set of equations consistently, we require additional four constraints. These are called \SSC~and they are widely studied in the literature. Here we briefly introduce some of these conditions and describe their important features: 
\begin{itemize}

\item $\textit{The Papapetrou and Corinaldesi condition}$, $S^{0i}=0$ \cite{corinaldesi2003spinning}. This would simply imply that there is no dipolar mass moment, \textit{i.e.} the center of mass in the particle's frame coincides with the observed center of mass in the chosen frame. 
\item $\textit{The Mathisson-Pirani supplementary condition}$, $S^{a b}U_b = 0$ \cite{Pirani:1956tn}. This condition is well studied upto some extend, and the predicted orbits are with helical structure. Initially it was believed to be unphysical, while recently it is has been shown that they have a physical interpretation \cite{Costa:2011zn,Costa:2017kdr}. The rest mass with respect to $U^a$ is given as, $m=-P^aU_a=$ constant.
\item $\textit{The Tulczyjew-Dixon supplementary condition}$, $S^{a b}P_{b} =0 $ \cite{tulczyjew1959motion}, is extensively studied in several works \cite{Kyrian:2007zz,Hartl:2003da,Hackmann:2014tga,Han:2008zzf}. This gives an exact physical solution of \MP~equations. In this case the dynamical mass, $\mu=\sqrt{-P^aP_a}$ is conserved.
\item $\textit{The Newton-Wigner condition}$, $S^{a b}\omega_{b}=0$ \cite{Newton:1949cq}, gives a Hamiltonian approach to the motion of spinning particles \cite{Barausse:2009aa}. The $\omega^{a}$ is given as, $\omega^{a}=P^a/\mu+ \phi^b$, where $\phi^b$ is a timelike vector. This would help to improve the phenomenological approach to understand gravitational waves dynamics. At the same time, it neither conserves total spin nor mass of the test particle.
\end{itemize}
A detailed discussion on various \SSC s and their connections to internal properties of the spinning particles can be found in Ref.~\cite{Lukes-Gerakopoulos:2014dma}. However in the present context, we start with the \T~or \P~constraint and investigate various possibilities of circular orbits at constant altitudes. Following this, we shall study the similar situations with \NW~\SSC~and compare the respective results.


\subsection{\T~or \P~\SSC}
In this section, we shall briefly discuss the evolution equations for a spinning particle in the Kerr spacetime within the framework of linear spin approximation.  We  explicitly use the \T~or \P~condition which are the same in this limit. The difference between  the four momentum and velocity  are of  higher order in $S$, {\it i.e.}, $\mathcal{O}(S^2)$.  With these conditions, the \MP~equations  simplify to~\cite{apostolatos1996spinning}:
\begin{equation}
\dfrac{DU^k}{d\tau} = -\dfrac{1}{2 m}R^k_{~b c d}U^b S^{c d}+\mathcal{O}(S^2) , \qquad \dfrac{DS^{ab}}{d\tau} =  0+\mathcal{O}(S^2), \quad \text{and} \quad m = \mu +\mathcal{O}(S^2).
\label{Eq_Linear_Appx}
\end{equation}
Now for computational convenience we use the standard spin four vector $S^a$ instead of the spin tensor, $S^{ab}$ given by:
\begin{equation}
S^{a} = \dfrac{\epsilon^{a b c d}}{2\sqrt{-g}}U_b S_{c d}, \qquad S^{ab}= \dfrac{1}{\sqrt{-g}}\epsilon^{a b c d}U_{c}S_{d}.
\label{eq:Spin_Vector}
\end{equation}
Where \enquote*{$g$} is the determinant of the metric and is always negative. The explicit form of the metric tensor in Boyer-Lindquist coordinates $(t,r,\theta,\phi)$ is given as
\begin{equation}
ds^2=-\dfrac{\Delta}{\Sigma} \Big(dt-a \sin^2\theta d\phi \Big)^2 +\dfrac{\Sigma}{\Delta}dr^2+\Sigma d\theta^2+\dfrac{\sin^2\theta}{\Sigma} \Big(-a dt+(r^2+a^2)d\phi\Big)^2,
\end{equation}
Here, \enquote*{$\Delta$} and \enquote*{$\Sigma$} has usual meanings, {\it i.e.}, $\Delta = r^2-2 M r+a^2$ and $\Sigma = r^2+a^2 \cos^2\theta$. For further analysis, we use the tetrad formalism \cite{Saijo:1998mn} with the components of the tetrads are given by
\begin{eqnarray}
e^{(0)}_{\mu} & = & \left(\sqrt{\dfrac{\Delta}{\Sigma}},0,0,-a \sin^2\theta \sqrt{\dfrac{\Delta}{\Sigma}} \right), \nonumber \\
e^{(1)}_{\mu} & = & \left(0,\sqrt{\dfrac{\Sigma}{\Delta}},0,0\right) \nonumber, \\
e^{(2)}_{\mu} & = & \left(0,0,\sqrt{\Sigma},0 \right) \nonumber, \\
e^{(3)}_{\mu} & = & \left(\dfrac{-a \sin\theta}{\sqrt{\Sigma}},0,0,\dfrac{r^2+a^2}{\sqrt{\Sigma}}\sin\theta \right).
\label{eq:Tetrad}
\end{eqnarray}
 The inverse of the tetrad given in \ref{eq:Tetrad} can be easily computed with the relation:
\begin{equation}
e^{a}_{(\mu)}=\eta_{(\mu)(\nu)}g^{a b}e^{(\nu)}_{~b}.
\label{eq:Inverse_Tetrad}
\end{equation} 
Where, $\eta_{(\mu) (\nu)}$ is given as,

\[\begin{bmatrix}
-1 & 0 & 0 & 0 \\
 0 & 1 & 0 & 0 \\
 0 & 0 & 1 & 0 \\
 0 & 0 & 0 & 1
\end{bmatrix}\]
\noindent
For a complete description of the evolution of the system, we start with a spinning particle with the spin vector of the form $S \equiv(S^t,S^r,S^{\theta},S^{\phi})$, moving in a circular orbit ($\dot{r}=\ddot{r} = 0$) at a constant altitude ($\dot{\theta}=\ddot{\theta}=0$). 
Following the \T~supplementary condition, we get
\begin{equation}
 S^{(0)}U^{(0)}-S^{(3)}U^{(3)}=0.
 \label{eq:Spin_constraint}
\end{equation}
where \enquote*{()} indicates any projection on the tetrad frame. Using \ref{Eq_Linear_Appx} and the conditions for circular orbits, it is easy to establish that both the time and $\phi$ components of the acceleration would identically vanish, {\it i.e}, $DU^{(0)}/d\tau=0=DU^{(3)}/d\tau$. Furthermore, from the R.H.S of \ref{Eq_Linear_Appx}, we must have $S^{(1)(2)}=S^{(3)}U^{(0)}-S^{(0)}U^{(3)} =0$ which contradicts the constraint given in \ref{eq:Spin_constraint}. Taking both these factors into account, we have two possibilities, either $S^{(0)}=S^{(3)}$  or $S^{(0)}=S^{(3)}=0$. For the first case we get, $\overline{\Omega}=U^{(3)}/U^{(0)}=1$, which is only possible for a light-like trajectory. Thus for time-like  orbits, we could only have $S^{(0)}=S^{(3)}=0$ and the resulting spin vector follows $S \equiv (0,S^{r},S^{\theta},0)$. In this case, the spin three vector would not be parallel or anti-parallel to the rotational axis of the black hole, instead has a nonzero inclination with respect to it.

Furthermore, the nonzero components correspond to the radial and angular equations can be reduced to:
\begin{eqnarray}
\Lambda_1 + \Lambda_2 \overline{\Omega}^2 + 2 \Lambda_3 \overline{\Omega} & = & - e^1_{(1)}\left[3 R_{(1)(3)(1)(3)}~\overline{\Omega}~S^{(2)}+R_{(1)(3)(0)(2)}~S^{(1)}~\left(1+\overline{\Omega}^2\right)\right], \label{eq:r_Eq}\\
 \tilde{\Lambda}_1 + \tilde{\Lambda}_2 \overline{\Omega}^2 + 2 \tilde{\Lambda}_3 \overline{\Omega} & = &  -e^2_{(2)}\left[3 R_{(1)(3)(1)(3)}~\overline{\Omega}~S^{(1)}-R_{(1)(3)(0)(2)}~S^{(2)}~\left(1+\overline{\Omega}^2\right)\right],\label{eq:theta_Eq}
 \end{eqnarray}
 with,
 \begin{eqnarray}
 \Lambda_1 &=& \Gamma^{1}_{3 3} \left(e^3_{(0)}\right)^2 + \Gamma^{1}_{0 0} \left(e^0_{(0)}\right)^2+2\Gamma^{1}_{0 3} \left(e^3_{(0)} e^0_{(0)}\right),\nonumber \\
  \Lambda_2 &=& \Gamma^{1}_{3 3} \left(e^3_{(3)}\right)^2 + \Gamma^{1}_{0 0} \left(e^0_{(3)}\right)^2+2\Gamma^{1}_{0 3} \left(e^3_{(3)} e^0_{(3)}\right),\nonumber \\
   \Lambda_3 &=& \Gamma^{1}_{3 3} \left(e^{3}_{(0)} e^3_{(3)}\right) + \Gamma^{1}_{0 0} \left(e^0_{(0)} e^0_{(3)}\right)+\Gamma^{1}_{0 3} \left(e^3_{(0)} e^0_{(3)}+ e^3_{(3)} e^0_{(0)}\right),\nonumber \\
 \tilde{ \Lambda}_1 &=& \Gamma^{2}_{3 3} \left(e^3_{(0)}\right)^2 + \Gamma^{2}_{0 0} \left(e^0_{(0)}\right)^2+2\Gamma^{2}_{0 3} \left(e^3_{(0)} e^0_{(0)}\right),\nonumber \\
  \tilde{\Lambda}_2 &=& \Gamma^{2}_{3 3} \left(e^3_{(3)}\right)^2 + \Gamma^{2}_{0 0} \left(e^0_{(3)}\right)^2+2\Gamma^{2}_{0 3} \left(e^3_{(3)} e^0_{(3)}\right),\nonumber \\
   \tilde{\Lambda}_3 &=& \Gamma^{2}_{3 3} \left(e^{3}_{(0)} e^3_{(3)}\right) + \Gamma^{2}_{0 0} \left(e^0_{(0)} e^0_{(3)}\right)+\Gamma^{2}_{0 3} \left(e^3_{(0)} e^0_{(3)}+ e^3_{(3)} e^0_{(0)}\right).
  \end{eqnarray}
 Where, $\Gamma$'s are the Christoffel symbols, $S^{(1)}$ and $S^{(2)}$ are the projection of radial ($S^r$) and angular ($S^{\theta}$) spin components respectively on the tetrad fame and we define $\overline{\Omega}=\dfrac{U^{(3)}}{U^{(0)}}$ to be the angular velocity of the particle in the tetrad frame. These equations describe the circular motion with only one parameter, $\overline{\Omega}$. The extremal values of $\overline{\Omega}$ are bounded by the angular velocity of photons, $\overline{\Omega}_{\rm ph}=\pm 1$. Now in principle one can solve \ref{eq:r_Eq} for $\overline{\Omega}$ and substitute in \ref{eq:theta_Eq}, and get relation between $r$ and $\theta$ for different spin values,
\begin{eqnarray}
\overline{\Omega} &=&\overline{\Omega}(r,\theta ,S^{(1)},S^{(2)},a),  \nonumber \\
\theta &=& \theta~(r,S^{(1)},S^{(2)},a).
\label{eq:R_Theta}
\end{eqnarray} 
We use the numerical approach to solve these equations in the next section. 
 
\subsection{Newton-Wigner \SSC}
In this section, we discuss the Newton Wigner formalism in detail and explicitly use it in the  framework of linear spin approximation while the background geometry is described by a Kerr black hole. As already described earlier, we define a timelike vector $\omega^a$ such that,
\begin{equation}
\omega^a=P^a/\mu+\phi^a.
\end{equation}
where, $\phi^a$ is an unit timelike vector. In this case, the Newton-Wigner constraint can be written in terms of $\omega^a$ as:
\begin{equation}
S^{a b}\omega_b=S^{a b}(\nu_b+\phi_b)=0.
\end{equation} 
With $\nu^b$ defines as the normalized momenta, $P^b=\mu \nu^b$. It should noted that neither mass ($\mu$ or $m$) nor the total spin is conserved in this formalism, while their differences appear only in the $\mathcal{O}(S^2)$. In the present context, neglecting terms containing $\mathcal{O}(S^2)$, we may consider them as conserved quantities. In addition, we constrain $\phi^a$ to satisfy,
\begin{equation}
\nu^a\nu_a=-1, \qquad \phi^a\phi_a=-1, \qquad \text{and} \qquad \nu^a\phi_a=-k.
\label{eq:Newton_Wigner_1}
\end{equation}  
Where, $k$ is a constant. We may now express the vectors in the chosen tetrad given in \ref{eq:Tetrad},
\begin{eqnarray}
\nu^{a} &=& a_0 e^a_{(0)}+a_3 e^a_{(3)},\nonumber \\
\phi^{a} &=& c_0 e^a_{(0)}+c_3 e^a_{(3)}.
\label{eq:Newton_Wigner_2}
\end{eqnarray} 
Using \ref{eq:Newton_Wigner_1} along with \ref{eq:Newton_Wigner_2}, we obtain,
\begin{eqnarray}
c_0 &=& a_0 k +a_3 \sqrt{k^2-1}, \nonumber\\
c_3 &=& a_3 k +a_0 \sqrt{k^2-1}. 
\label{eq:NW_Constraints}
\end{eqnarray}
As in the previous case, here also we introduce a spin four vector to simplify calculations. However, in this case, the vector need to be defined with respect to $\omega^a$ instead of four momentum ($P^a$) or velocity ($U^a$): 
\begin{equation}
S^{ab}=\dfrac{\epsilon^{abcd}\omega_cS_d}{\sqrt{-g}(-\omega^ m \omega_m)}=\dfrac{\epsilon^{abcd}\omega_cS_d}{\sqrt{-g}(1+k)}.
\label{eq:Spin_NW}
\end{equation}
The denominator of the above equation would contribute a factor of 2 which has been  incorporated in the spin vector.  The special case, $k=1$ corresponds to a trivial case with $c_0=a_0$ and $c_3=a_3$. In the present case, $\omega^{a}$  plays the same role  as velocity or momentum does in \T~condition. As evident from \ref{eq:Newton_Wigner_2}, $\omega^a$ would  have the specific form, $\omega \equiv (0,\omega^r,\omega^{\theta},0)$. Hereby, the spin vector would obey the same structure as in \T~\SSC~following the identical reason given there.
From \ref{eq:NW_Constraints} and \ref{eq:Spin_NW}, we can write the orbit equations as
\begin{eqnarray}
 \Lambda_1 + \Lambda_2 \overline{\Omega}^2_{\rm NW} + 2 \Lambda_3 \overline{\Omega}_{\rm NW}&=&  -e^1_{(1)}\left[ R_{(1)(3)(1)(3)} \left(3\overline{\Omega}_{\rm NW}+\alpha_1 \right) S^{(2)}+R_{(1)(3)(0)(2)}~S^{(1)}~\left(1+\overline{\Omega}^2_{\rm NW}+ \beta \right)\right], \nonumber \label{eq:r_Eq_NW}\\
 \tilde{\Lambda}_1 + \tilde{\Lambda}_2 \overline{\Omega}^2_{\rm NW} + 2 \tilde{\Lambda}_3 \overline{\Omega}_{\rm NW}&=& - e^2_{(2)}\left[ R_{(1)(3)(1)(3)} (3\overline{\Omega}_{\rm NW}+\alpha_2) S^{(1)}-R_{(1)(3)(0)(2)}S^{(2)}\left(1+\overline{\Omega}^2_{\rm NW}+\beta\right)\right].\nonumber
 \\
 \label{eq:theta_Eq_NW}
\end{eqnarray}
The $\alpha$'s and $\beta$ are given as:
\begin{equation}
\alpha_1 =\sqrt{\dfrac{k-1}{k+1}}\left\{2+\overline{\Omega}^2_{\rm NW}\right\}, \qquad 
\alpha_2 = \sqrt{\dfrac{k-1}{k+1}}\left\{1+2 \overline{\Omega}^2_{\rm NW}\right\}, \quad \text{and} \quad \beta= 2  \overline{\Omega}_{NW}\sqrt{\dfrac{k-1}{k+1}}.
\end{equation}
As one can see, the general dependence of these equations on $k$ is weak as the prefactor goes as, $\sqrt{\dfrac{k-1}{k+1}}$ and for a large value of $k$ it is close to unity. Hence, the orbit equations in both these formalism would differ by a small amount. The equations correspond to \T~or \P~\SSC~can be easily obtained by setting $\alpha_1$, $\alpha_2$ and $\beta$ to zero. Now we may rewrite the \ref{eq:theta_Eq_NW} similar to the previous case as in \ref{eq:R_Theta} and numerically solve them to compute the non-equatorial circular orbits: 
\begin{eqnarray}
\overline{\Omega}_{\rm NW} &=&\overline{\Omega}_{\rm NW}(r,\theta ,S^{(1)},S^{(2)},a),  \nonumber \\
\theta &=& \theta~(r,S^{(1)},S^{(2)},a).
\label{eq:R_Theta_NW}
\end{eqnarray} 
Similar to the previous case, we shall solve these equations in the next section.
\subsection{Conserved quantities : energy, momentum and Carter constant}
The Killing vectors are closely connected with the integrals of motion. For a geodesic motion and the scalar product of four-momentum with Killing vector is conserved. 
In the case of spinning particles, the conserved quantities get modified depending on the spin of the particle. For a killing vector field $\mathcal{K}^a$, the corresponding conserved quantity is written as \cite{rudiger1981conserved,rudiger1983conserved}:
\begin{equation}
\mathcal{C}=\mathcal{K}^aP_a-\dfrac{1}{2} S^{a b}\mathcal{K}_{a;b}.
\label{Eq:Conserved_quantity}
\end{equation} 
Where the semicolon ($;$) is defined as the covariant derivative. As the Kerr spacetime has two Killing vectors, a timelike ($\xi^a$) and a spacelike ($\eta^a$), the corresponding conserved quantities are given by:
\begin{equation}
E=-\mathcal{C}_t=-\xi^aP_a+\dfrac{1}{2} S^{a b}{\xi}_{a;b}, \qquad \text{and} \qquad 
J_z=\mathcal{C}_{\phi}=\eta^aP_a-\dfrac{1}{2} S^{a b}{\eta}_{a;b}.
\end{equation}
Unlike the geodesics, neither energy $(-\xi^aP_a)$ nor the angular momentum $(\eta^aP_a)$ is conserved in case of a spinning particle. Instead, we have the conserved quantities are, $-\mathcal{C}_t$ and $\mathcal{C}_{\phi}$ and they become energy and angular momentum only for specific case $S=0$.

In addition to the above conserved quantities, there is another constant of motion related to total angular momentum of a particle. This is called Carter constant \cite{Carter:1968rr,Carter:1979fe,dewitt1973black}. A general prescription to define total angular momentum is more involved in general relativity. In fact, it did not receive much attention until Carter came up with this non-trivial constant to describe the geodesic motion in a Kerr black hole. It turns out that this constant is closely related to the total angular momentum of a particle and for a static spacetime it is exactly same as the square of total angular momentum \cite{Rosquist:2007uw,Will:2008ys,Mukherjee:2015oaa}. Presence of this constant makes the trajectories completely integrable in the Kerr spacetime. Even though, so far there is no general notion of Carter like constant in case of spinning particles, one can establish an approximate formula for Carter constant valid only up to linear order in spin \cite{gibbons1993susy}. This formula is evaluated in the framework of Einstein's theory of general relativity. It is explicitly used by Tanaka \textit{et. al.} to demonstrate that the adiabatic approximation can be applicable in case of spinning particle upto linear order on the equatorial plane \cite{Tanaka:1996ht}. This is given by:
\begin{eqnarray}
\dfrac{Q}{m^2} &=& \left\{ \left(\Sigma \Big[(U^{(0)})^2-(U^{(1)})^2\Big]-r^2\right) \right \}-\dfrac{2 a\sin\theta}{\sqrt{\Sigma}}\biggl\{r \left(U^{(0)}S^{(1)(3)}-2 U^{(3)}S^{(1)(0)}+U^{(1)}S^{(3)(0)}\right)+\nonumber \\
& & a \cos\theta U^{(3)}S^{(2)(3)}\biggr\}- \dfrac{2 \sqrt{\Delta}}{\sqrt{\Sigma}}\left\{a \cos\theta \left(2 U^{(0)}S^{(2)(3)}-U^{(3)}S^{(2)(0)}+U^{(2)}S^{(3)(0)}\right)-r U^{(0)}S^{(1)(0)}\right\}. \nonumber \\
\label{eq:Carter_Spinning}
\end{eqnarray}
Before dealing with the more general case of a rotating black hole, we first investigate the properties of the above constant in a Schwarzschild black hole. By setting $a=0$, \ref{eq:Carter_Spinning} become,
\begin{equation}
\dfrac{Q}{m^2}=r^2 \left\{ \left( U^{(2)}\right)^2+\left(U^{(3)}\right)^2 \right\} +2 \sqrt{\Delta} U^{(0)} S^{(1)(0)}.
\end{equation}  
The above equation can be further simplified to a familiar form by using the explicit forms of the tetrads suggesting the first term to be the square of total angular momentum ($L$), while the second term can be written in terms of the spin vector.
\begin{equation}
r^2 \left\{ \left( U^{(2)}\right)^2+\left(U^{(3)}\right)^2 \right\}=\dfrac{L^2}{m^2} = (U_{\theta})^2+\dfrac{(U_{\phi})^2}{\sin^2\theta}. 
\end{equation}
So we may conclude,
\begin{equation}
Q=L^2+2 m^2 \Delta r \sin\theta U^t U^{\phi}S^{\theta}. 
\label{eq:Carter_Spinning_2}
\end{equation}
It is interesting to see that the extra term is proportional to the spin vector ($S^{\theta}$) and for a limit $S \rightarrow 0$, $Q \rightarrow L^2$. Now we compute the total angular momentum (orbital+spin) of a spinning test particle and explicitly show this matches with \ref{eq:Carter_Spinning_2}.

We have already discussed how a killing vector is useful to exploit various symmetries in a geometry. Unlike a Kerr black hole, Schwarzschild spacetime is endowed with spherical symmetry and contain three spacelike killing vectors :
\begin{equation}
\eta^a_1 = x\dfrac{\partial}{\partial y}-y\dfrac{\partial}{\partial x}, \qquad \eta^a_2 = y\dfrac{\partial}{\partial z}-z\dfrac{\partial}{\partial y}, \qquad \text{and} \qquad \eta^a_3 = z\dfrac{\partial}{\partial x}-x\dfrac{\partial}{\partial z}.
\end{equation}
Now from \ref{Eq:Conserved_quantity}, we write down the each conserved quantities explicitly:
\begin{eqnarray}
\dfrac{J_x}{m} &=& \left\{-\sin\phi U_{\theta}-\cot \theta \cos \phi U_{\phi}\right\}+ (S^r U^t-S^t U^r)\left(\dfrac{\cos \phi \cos 2\theta}{2 \sin\theta}-\dfrac{\cos \phi }{2 \sin \theta }\right)- \nonumber\\
& & r \sin\theta \sin\phi (1-2 M/r)(S^t U^{\phi}-S^{\phi}U^t)-r \cos\theta \cos\phi (1-2M/r)(S^{\theta}U^t-S^t U^{\theta}), \nonumber \\
\dfrac{J_y}{m} &=& \left\{\cos\phi U_{\theta}-\cot \theta \sin \phi U_{\phi}\right\}+ (S^r U^t-S^t U^r)\left(\dfrac{\sin \phi \cos 2\theta}{2 \sin\theta}-\dfrac{\sin \phi }{2 \sin \theta }\right)+ \nonumber\\
& & r \sin\theta \cos\phi (1-2 M/r)(S^t U^{\phi}-S^{\phi}U^t)-r \cos\theta \sin\phi (1-2M/r)(S^{\theta}U^t-S^t U^{\theta}),\nonumber \\
\dfrac{J_z}{m} &=& U_{\phi}+(r-2 M)\sin\theta \left(U^t S^{\theta}-U^{\theta}S^t\right)+\cos\theta \left(S^t U^r-S^r U^t\right).
\end{eqnarray}
With the above equations, it is easy to show that
\begin{equation}
J^2 =J^2_x+J^2_y+J^2_z=L^2+2 m^2 \Delta r \sin\theta U^t U^{\phi}S^{\theta} + \mathcal{O}(S^2) \approx Q.
\end{equation}
So as we claimed earlier, the Carter constant for spinning particle with the linear spin approximation is similar to the total angular momentum of the particle.

In case of a rotating geometry, we shall describe a new quantity as effective Carter constant, $K_s =Q-(J_z-a E)^2$. Note that, this quantity would vanish in case of a geodesic trajectory on the equatorial plane. In the present context, one can also compute $K_s$ for circular orbits of spinning particles lying close to the equatorial plane. With a series expansion around $\theta=\pi/2+\eta$, one can accommodate the terms linear in $\eta$,
\begin{equation}
K_s \approx 2 a S^{z} -\dfrac{2  S^{(1)}}{r^2 \sqrt{\Delta}} \left \{E J_z r^3-2 a M (J_z-a E)^2 \right\} \eta + \mathcal{O}(\eta^2).
\label{Eq:Appx_Carter_2}
\end{equation}
The first term is a coupling between the spin component of the particle with the angular 
momentum of the black hole, while the second term is related to the square of momenta. It is interesting to note that the second term is proportional to both $S^{(1)}$ and $\eta$. 
It should be noted that  the approximation $\theta=\dfrac{\pi}{2} +\eta$ is used only to demonstrate the properties of the $K_s$. However, the results deduced in the article are valid for any arbitrary angle $\theta$. 

Before closing this section, we would like to remind a significant departure of spinning particles from the non-spinning trajectories. For geodesic orbits, the Carter constant identically vanishes on the equatorial plane. However spinning particles with aligned spin, i.e, with $S^{(1)}$ to be zero, stay on the equatorial plane even if the Carter constant is nonzero there \cite{Sajal:Prep,Tanaka:1996ht}.

\section{Non-equatorial orbits: constraining r and theta ($\theta$)}\label{sec:R_Theta}
Before delving into the spinning particle, we first investigate the possibilities of circular geodesics on the non-equatorial plane of a Kerr black hole. We start with the effective potential in radial ($V_r$) and angular ($V_{\theta}$) direction \cite{chandrasekhar1998mathematical,o1995geometry}:
\begin{eqnarray}
V(\theta) & = & K-L_z^2 \cot^2\theta+ a^2 (E^2-m^2)\cos^2\theta, \nonumber \\
V(r)      & = & E^2 r^4-(L_z^2-a^2 E^2)r^2+2(L_z-a E)^2 r-(m^2 r^2+K)\Delta \, .
\end{eqnarray}
where $K$ is the effective Carter constant for a geodesic. The necessary and sufficient conditions for a circular orbit are given by, $V(r)=0 $ and $\dfrac{dV(r)}{dr}=0$. In addition to this, a circular orbit at constant altitude also has to satisfy $V(\theta)=0$ and $\dfrac{dV(\theta)}{d\theta}=0$,
\begin{equation}
\dfrac{dV(\theta)}{d\theta}=2 \cos\theta \left\{L_z^2 \csc^3\theta -a^2 (E^2-m^2)\sin\theta\right\}=0.
\end{equation} 
This immediately suggests, either $\theta=\pi/2$ or $L_z^2 = a^2 (E^2-m^2)\sin^4\theta$. In the first case with, $\theta=\pi/2$, it is easy to show that $V(\theta)$ vanishes only when $K=0$. Now one can employ the radial potential $V(r)$ to show that $K=0$ indeed describes a circular orbit on the equatorial plane. On the other hand, for $L_z^2 = a^2 (E^2-m^2)\sin^4\theta$, bound circular orbits are unlikely to appear as they consists with $E^2-m^2<0$ and this is inconsistent with $L_z^2>0$. Hence, one may conclude that circular orbits for geodesic trajectories can only exist on the equatorial plane of a Kerr black hole \cite{de1979non}. However, if one relax the constraint of vanishing $\dot{\theta}$, i.e, $\dot{\theta} \neq 0$ , spherical orbits are likely to appear in Kerr background \cite{Wilkins:1972rs}.

The situation is quite different in the case of a spinning particle and non-equatorial circular orbits can be obtained from \MP~equations. Next we shall discuss these orbits for both the spin supplementary conditions.
\subsection{\T~or \P~\SSC  }
\begin{figure}[htp]
\subfloat[The co-rotating circular orbits are shown in the non-equatorial planes for different angular momentum of the black hole. For a large angular momentum of the black hole, the orbits are dragged close to the horizon. The spin components are fixed at $S^{(i)}=(0,-0.015M,-0.01M,0)$ while  the angular momentum \enquote*{a} is shown in the inset.  \label{fig:Non_Eq1}]{%
  \includegraphics[height=5.8cm,width=.49\linewidth]{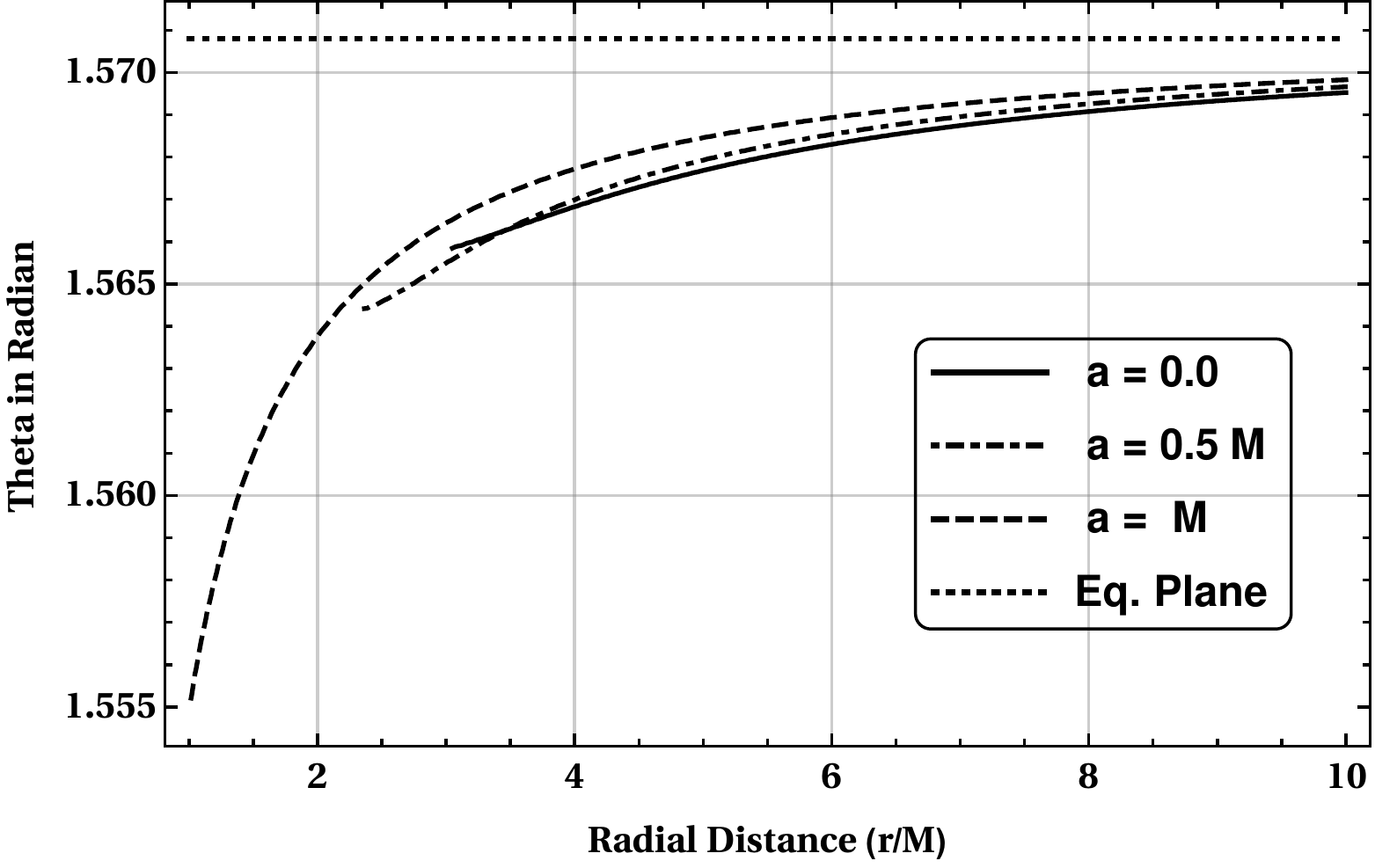}%
}\hfill
\subfloat[The counter-rotating circular orbits are shown for $S^{(i)}=(0,-0.015M,-0.01M,0)$. They move away from the horizon as one increases the value of the black hole's angular momentum. As the spinning particle moves close to the horizon, it gets more deviated from the equatorial plane. The  black hole's momenta are shown in the inset.\label{fig:Non_Eq2}]{%
  \includegraphics[height=5.8cm,width=.49\linewidth]{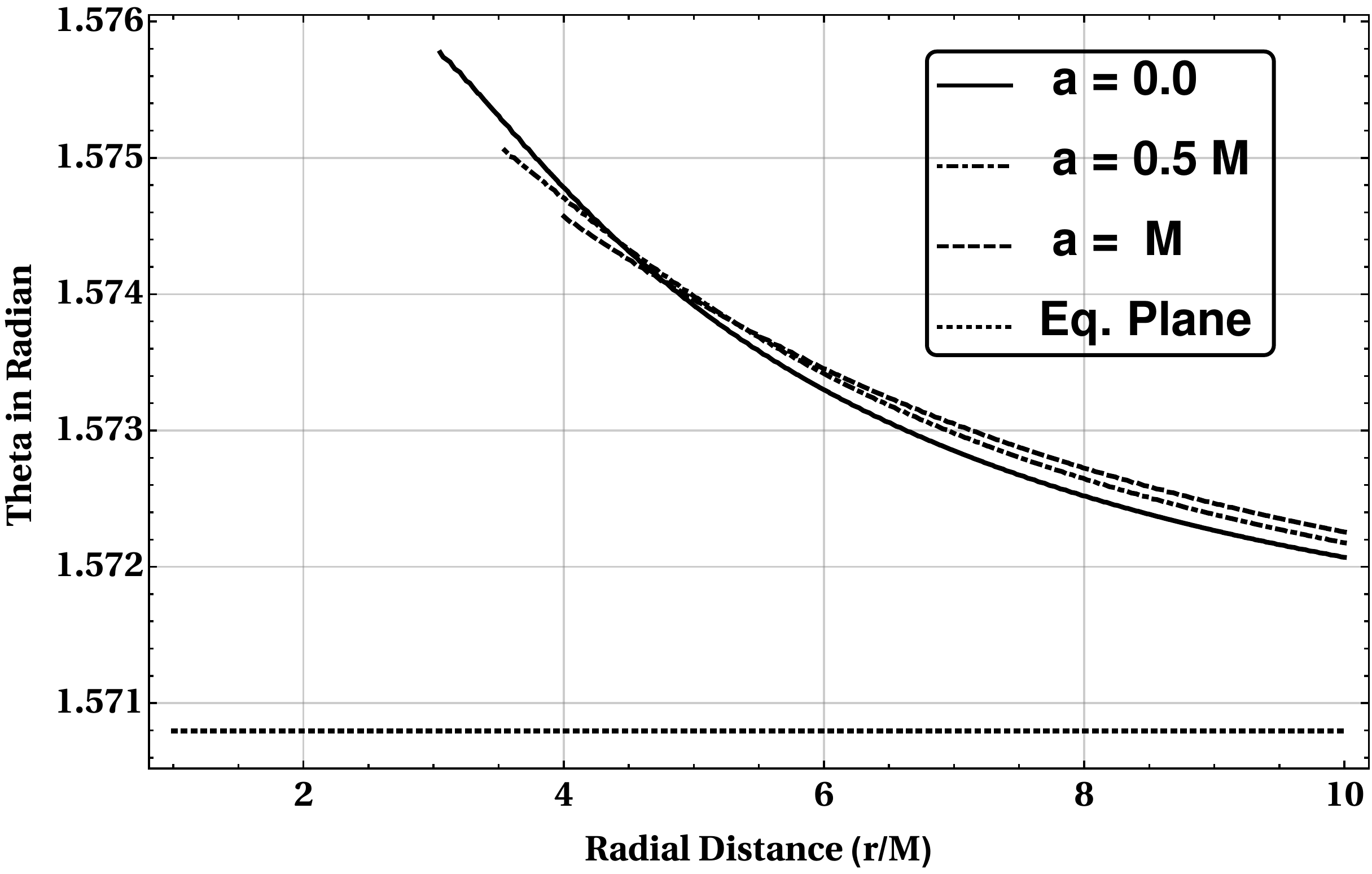}%
}
\caption{Non-equatorial circular orbits are shown for spinning particles in a Kerr spacetime. In addition, we would also like to point out that the innermost unstable circular orbits are slightly deviated from the geodesic limits due to the non-vanishing spin. For example, with $a=0.5M$, the innermost direct and retrograde unstable circular orbits for a geodesic trajectory exists at $r_{\rm direct}=2.3473M$ and $r_{\rm retro}=3.53209M$ respectively, while in our case it is shifted to $r^{s}_{\rm direct}=2.36403M$ and $r^{s}_{\rm retro}=3.52603M$. Similarly with $a=M$, the limits for a geodesic is given as $r_{\rm direct}=M$ and $r_{\rm retro}=4M$ for direct and retrograde respectively and in our study it becomes $r^s_{\rm direct}=1.00001 M$ and $r^{s}_{\rm retro}=4.00287M$. The limit with non-rotating case shifted to $r^s_{\rm direct}=3.04916M$ and $r^s_{\rm retro}=3.05118M$ which is close to $r_{\rm direct}=r_{\rm retro}=3M$ defining the innermost unstable circular orbit in Schwarzschild black hole.}
\label{fig:Non_Eq}
\end{figure}

\begin{figure}[htp]
\subfloat[The co-rotating circular orbits are shown for $S^{(2)}=-0.01 M$ while $S^{(1)}$ varies as shown in the inset.  \label{fig:Slope_01}]{%
  \includegraphics[height=5.8cm,width=.49\linewidth]{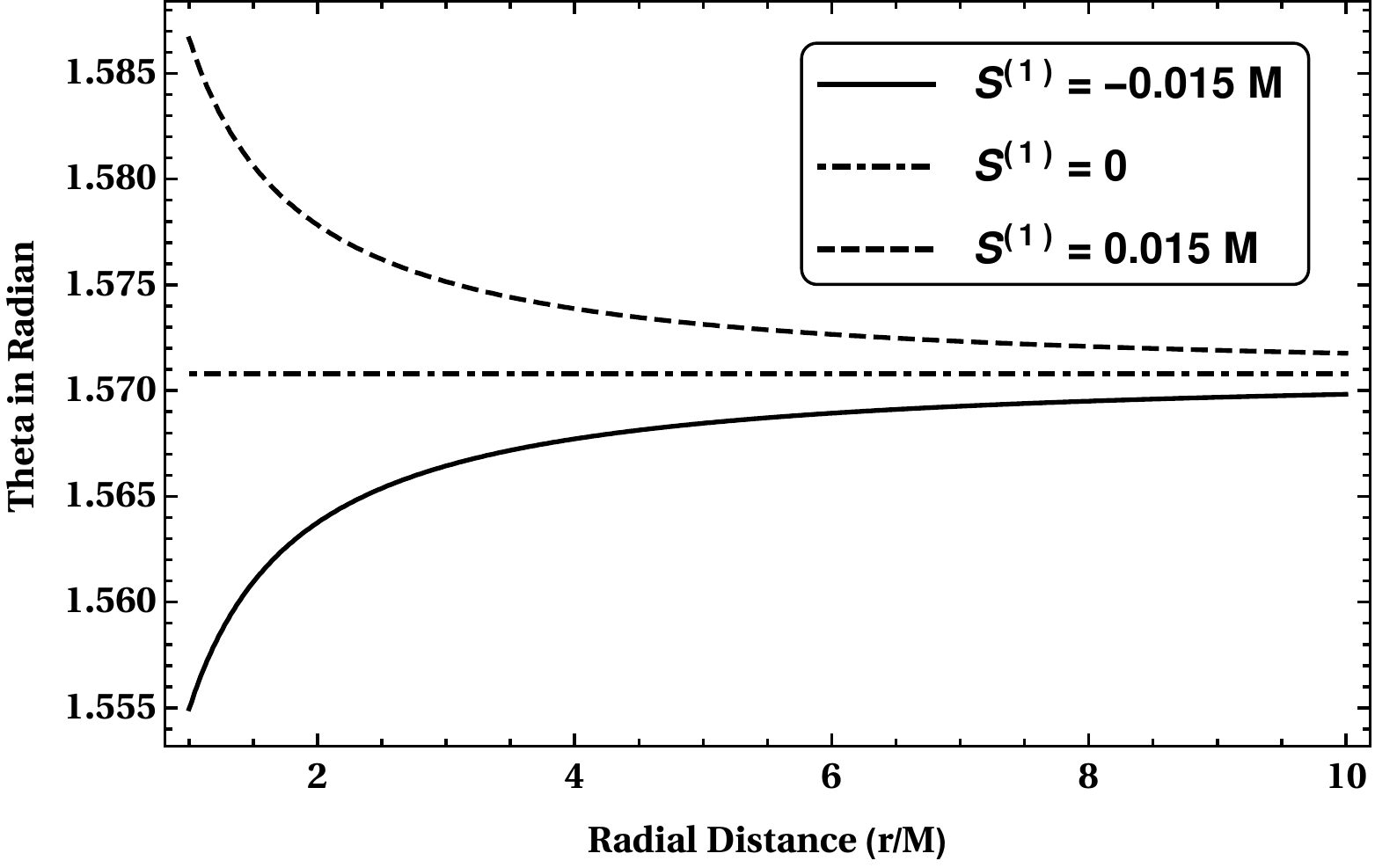}%
}\hfill
\subfloat[The counter-rotating circular orbits for a spinning particle are shown in the above figure, $S^{(2)}$ is fixed at $-0.01M$ while $S^{(1)}$ changes accordingly. \label{fig:Slope_02}]{%
  \includegraphics[height=5.8cm,width=.49\linewidth]{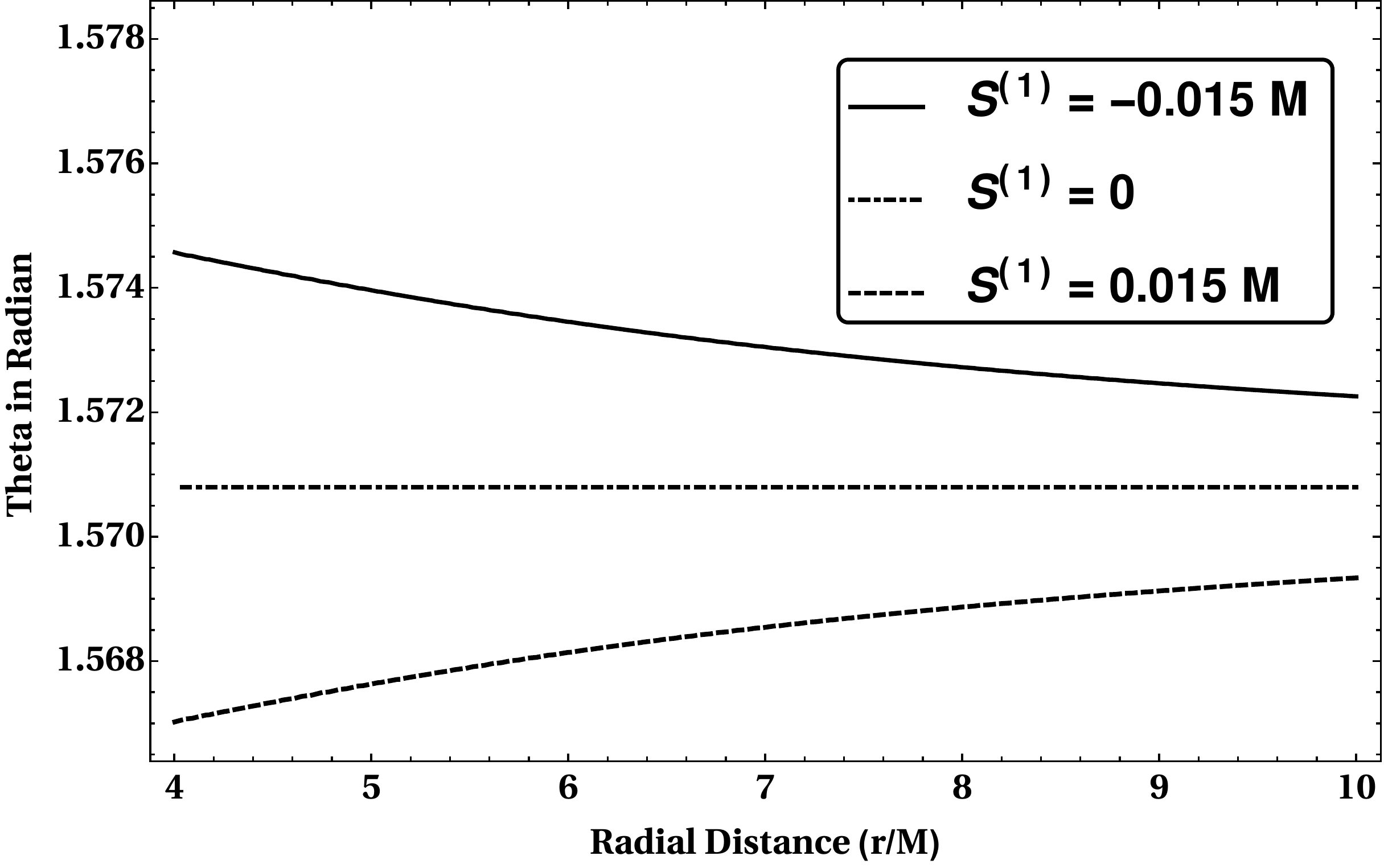}%
}
\caption{Non-equatorial circular orbits for a spinning are shown in a maximally rotating Kerr black hole. For a vanishing $S^{(1)}$, all the orbits reside in the equatorial plane.}
\label{fig:Slope}
\end{figure}

Here we numerically solve \ref{eq:R_Theta} and obtain ($r,\theta$) for a given value of the spin parameter. For each polar angle \enquote*{$\theta$}, there is only one possible radial coordinate \enquote*{r} that satisfies the equation of motion. The plot for \enquote*{$\theta$} as a function of \enquote*{r} is shown in \ref{fig:Non_Eq} for a given spin vector, $S^{(i)}=(0,\-0.015M,-0.01M,0)$ and angular momentum $a=\{0,0.5M,M\}$ for both co-rotating and counter-rotating orbits. These orbits behave as a small perturbation from the geodesic trajectories as they appear very close to the equatorial plane. It is shown that the deviation from the equatorial plane not only depends on the sign of the spin vector, but also on their direction of rotations. The counter rotating orbits shown in \ref{fig:Non_Eq2} cease to exist beyond $r \approx 4M$ for a maximally rotating Kerr black hole ($a=M$), while co-rotating orbits continue appear even close to the horizon as shown \ref{fig:Non_Eq1}. 
Let us now briefly discuss the dependence of orbital inclination on the spin vector. The corresponding slope for these orbits close to the equatorial plane can be computed using the orbit equation. For convenience, we set $a=M$ and then differentiate the equation with respect to r. Afterwards, we evaluate $d\theta/dr$ at $\theta=\pi/2$ and this is given as,
\begin{equation}
 \dfrac{d\theta}{dr} \approx-\dfrac{12 M S^{(1)} \overline{\Omega}}{M^2 r + 2 M r (r-M) \overline{\Omega} + ( r M^2 + r^3 ) \overline{\Omega}^2 + 6 M^{2} S^{(2)}(1 + \overline{\Omega}^2)}.
 \label{eq:slope}
 \end{equation}
For a co-rotating trajectory with $S^{(1)}<0$, the orbits get close to the $\theta = 0$ axis while the behavior is completely opposite for a counter-rotating orbit. In addition, the polar angle strongly depends on the spin component $S^{(1)}$ while has a weak dependency on $S^{(2)}$. By substituting $S^{(1)}$ to zero, the slope would identically vanish and the particle resides in the equatorial plane. It is depicted in \ref{fig:Slope}. However, with $S^{(2)}$ set to zero and $S^{(1)}$ to be nonzero, the particle will have non-equatorial trajectories.
\subsection{\NW~\SSC}
\begin{figure}[htp]
\subfloat[The above figure shows the co-rotating circular orbits in non-equatorial planes with the spin vector remain similar to the previous case, $S^{(i)} =(0,-0.015M,-0.01M,0)$ . \label{fig:Non_Eq1_NW}]{%
  \includegraphics[height=5.8cm,width=.49\linewidth]{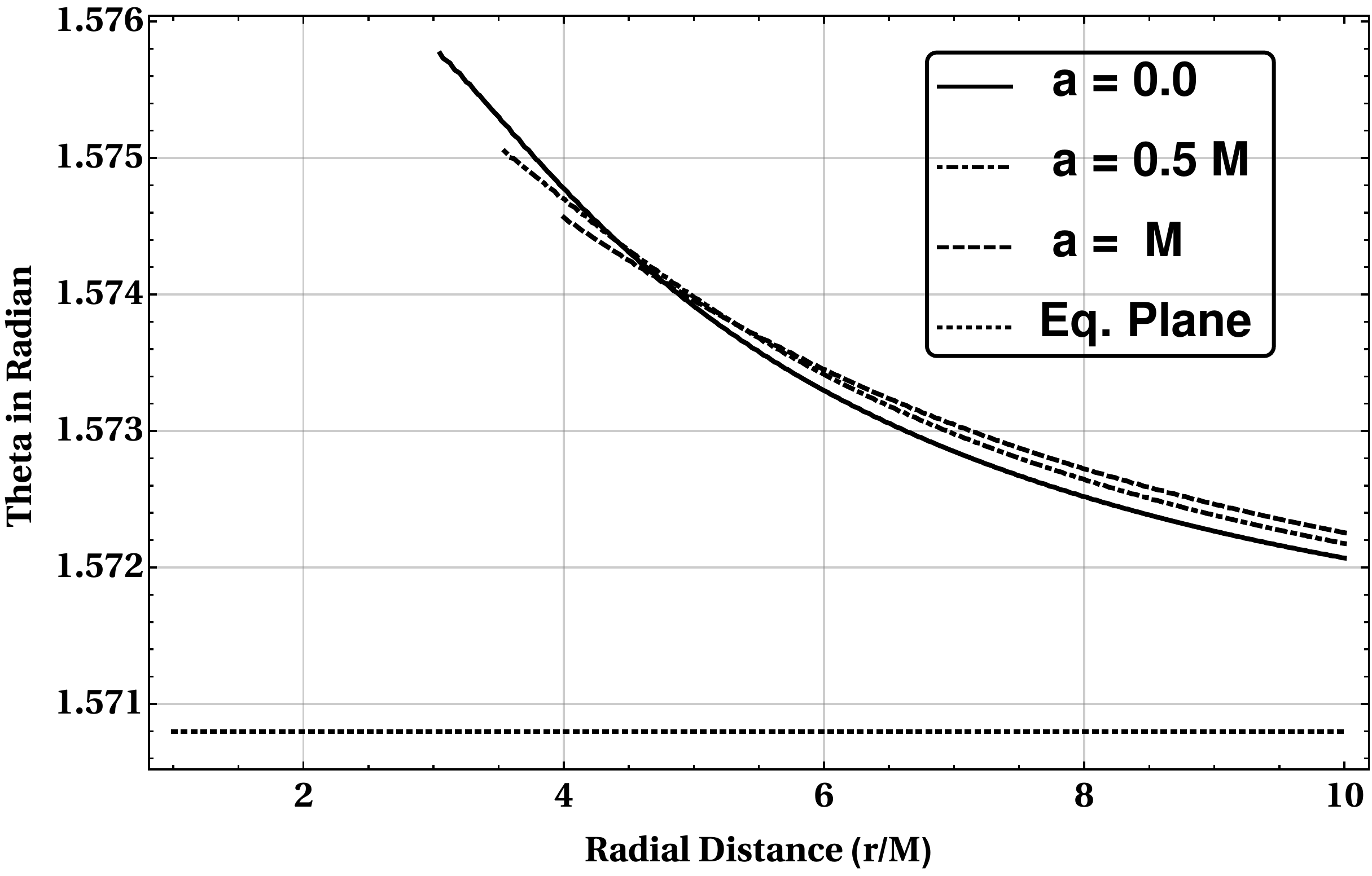}%
}\hfill
\subfloat[The counter-rotating circular orbits are shown with the \NW~condition.\label{fig:Non_Eq2_NW}]{%
  \includegraphics[height=5.8cm,width=.49\linewidth]{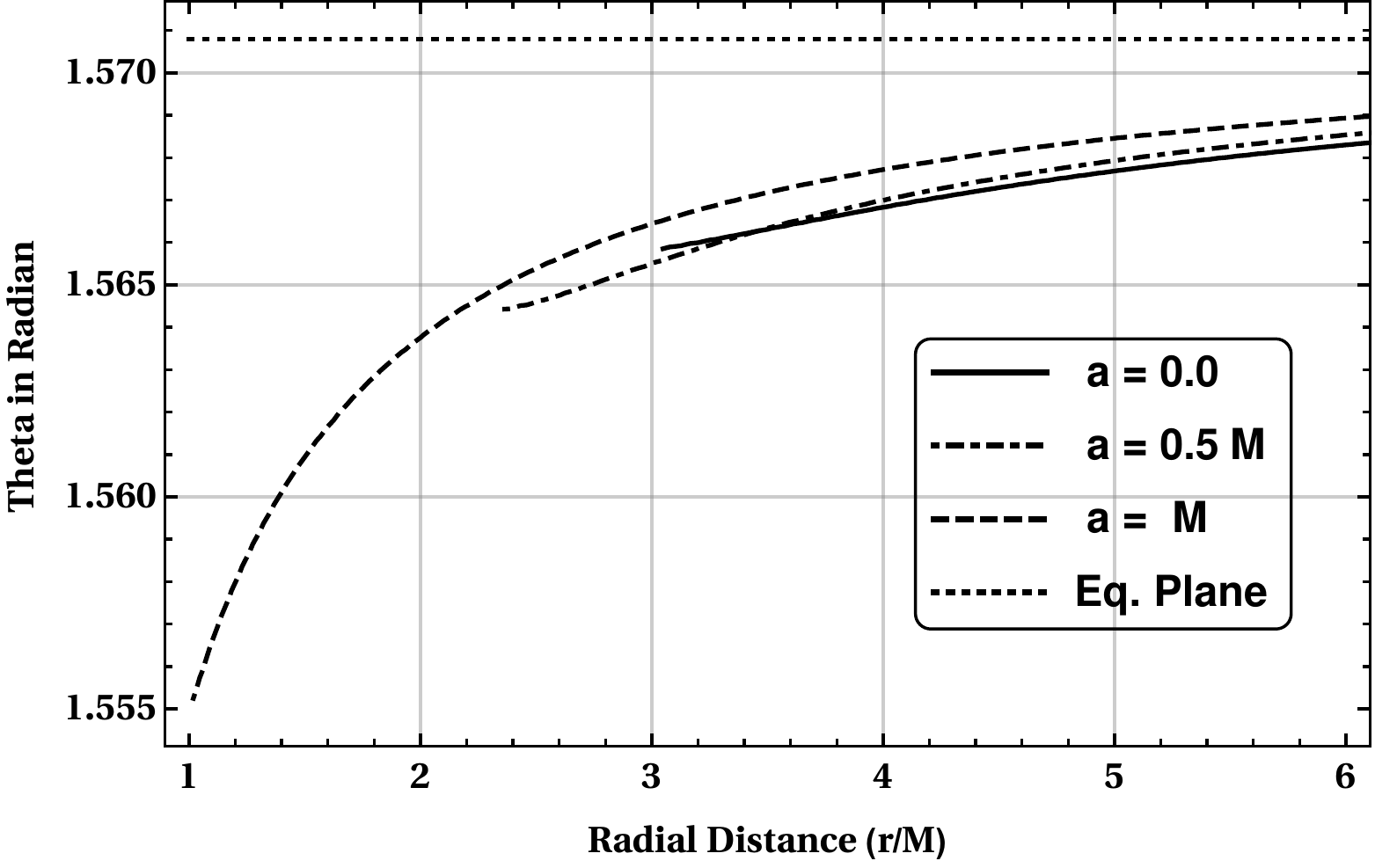}%
}
\caption{Non-equatorial circular orbits, both co-rotating and counter-rotating, are shown for spinning particles with \NW~constraint in a Kerr spacetime. The value of $k$ is fixed at 2. Similar to the previous case, one can show that the innermost unstable circular orbits are in fact within close proximity to the geodesic trajectories. For example, with $a=M$, the innermost unstable circular orbits for direct and retrograde directions are located at $r^s_{\rm direct}=1.01865M$ and $r^s_{\rm retro}=4.0024M$ respectively.}
\label{fig:Non_Eq_NW}
\end{figure}

\begin{figure}[htp]
\subfloat[ The above figure shows the co-rotating circular orbits for $S^{(i)}=(0,-0.015M,-0.01M,0)$ with two different \SSC s. Though the nature of the plots remain same as seen from \ref{fig:Non_Eq} and \ref{fig:Non_Eq_NW}, they differ in a small scale. \label{fig:Difference_01}]{%
  \includegraphics[height=5.8cm,width=.49\linewidth]{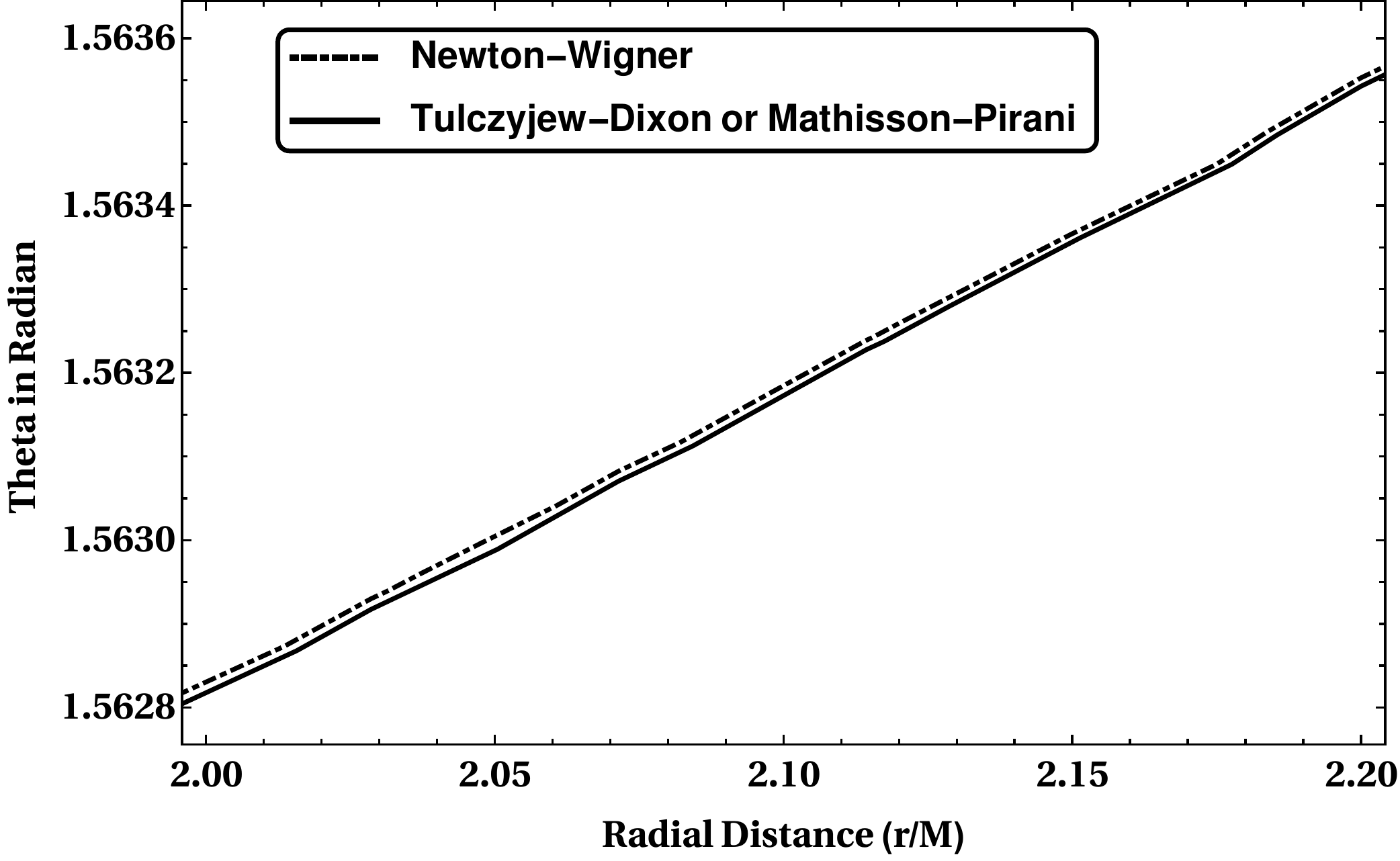}%
}\hfill
\subfloat[ The dependence of \MP~equations for different \SSC s~certainly increases with an increase of the spin parameters. A considerable ammount of difference is acheived with $S^{(i)}=(0,-0.05M,-0.045M,0)$. \label{fig:Difference_02}]{%
  \includegraphics[height=5.8cm,width=.49\linewidth]{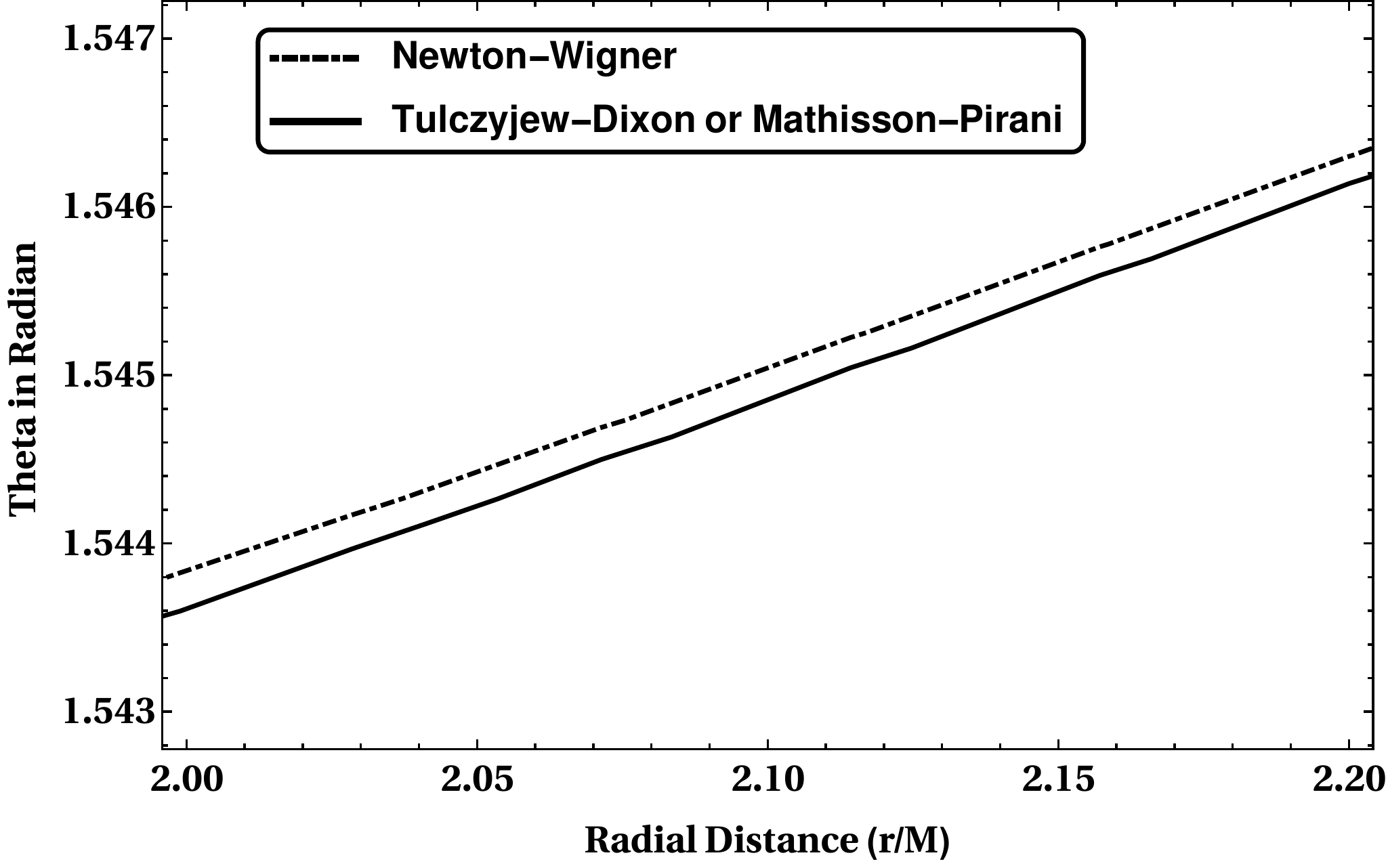}%
}
\caption{Figure shows a comparative study of non-equatorial orbits for different \SSC s~with the angular momentum of the black hole fixed at $a=0.8M$.}
\label{fig:difference}
\end{figure}

In the \NW~condition one has to solve \ref{eq:R_Theta_NW} along with \ref{eq:theta_Eq_NW} to compute the non-equatorial orbits, these are shown in \ref{fig:Non_Eq_NW}.
It should be noted that the dependence of these results on $k$ is weak and almost negligible. This is related to the prefactor $\sqrt{\dfrac{k-1}{k+1}}$ which has a maximum value of one as discussed earlier. The overall behavior is similar to the previous case, while the numerical values differ by a small amount, as shown in \ref{fig:Difference_01}. The difference between these two \SSC s~become significant only for larger values of spin as shown in \ref{fig:Difference_02}.

\begin{figure}[htp]
\centering
\includegraphics[scale=.6]{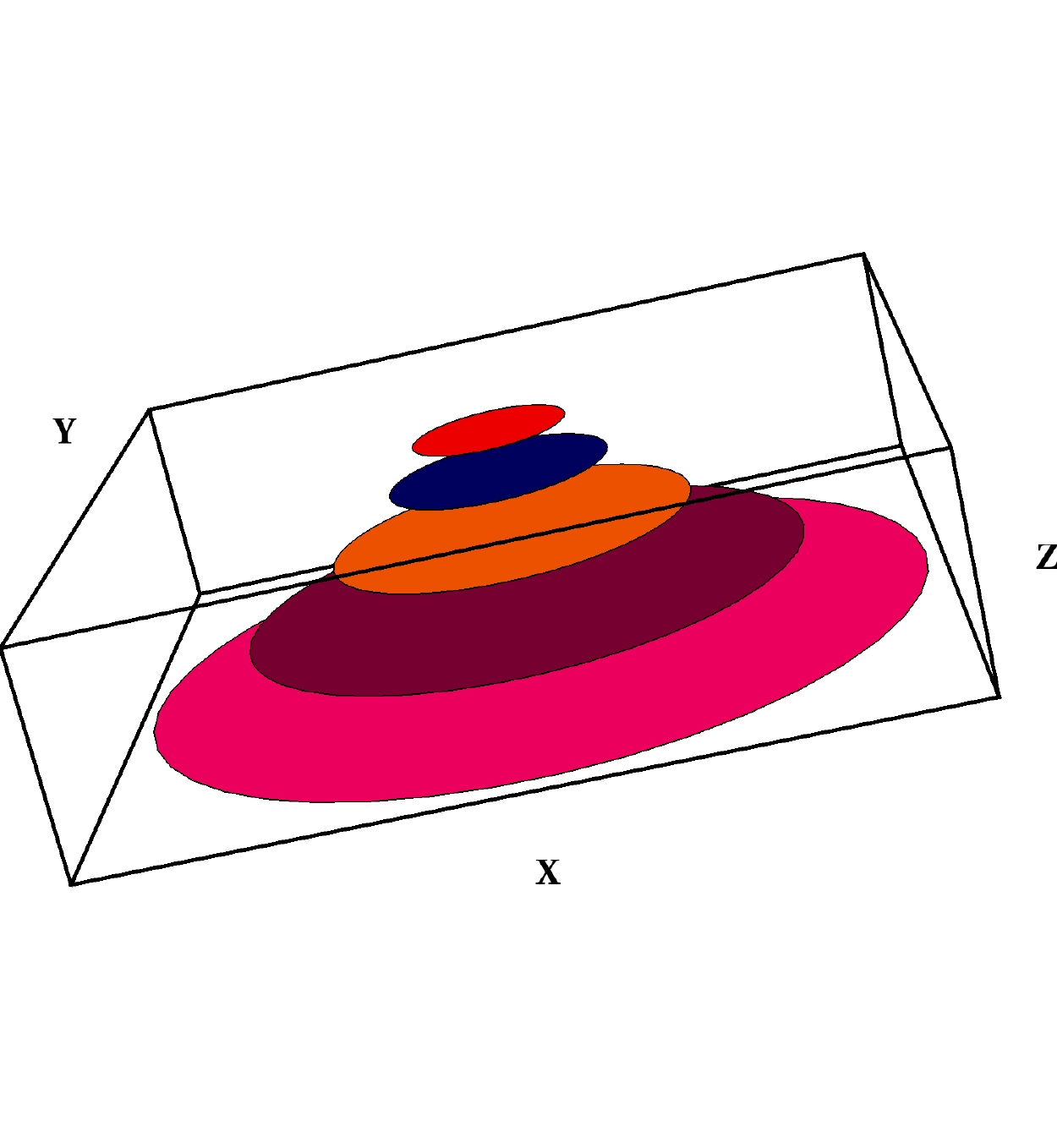}
\caption{The circular orbits are shown explicitly in the non-equatorial planes. The origin is located at $(0,0,r_s\cos\theta_s)$ while the radius is $R_s=r_s \sin\theta_s$. Scale of $r_s\cos\theta_s$ is raised by the square of logarithmic to realize the difference properly.}
\label{fig:3d}
\end{figure}

It is evident from \ref{fig:Non_Eq} and \ref{fig:Non_Eq_NW} that the radial coordinate ($r$) for each non-equatorial orbit is related to a specific value of angular coordinate ($\theta$). For a given spin value, if one choose to have a circular orbit at $\theta=\theta_s$, the corresponding radial coordinate takes a particular value of $r=r_s$. We can estimate the radius of such circular orbits at constant altitude as, $R_s=r_s\sin\theta_s$. As one gets closer to the horizon, it deviates furthermore from the equatorial plane, and also the radius ($R_s$) starts to decrease. This can be better explain in a graphical representation as shown in \ref{fig:3d}.
 \section{Circular orbits and stability analysis}\label{sec:Stability}
\begin{figure}[htp]
\centering
\includegraphics[scale=.5]{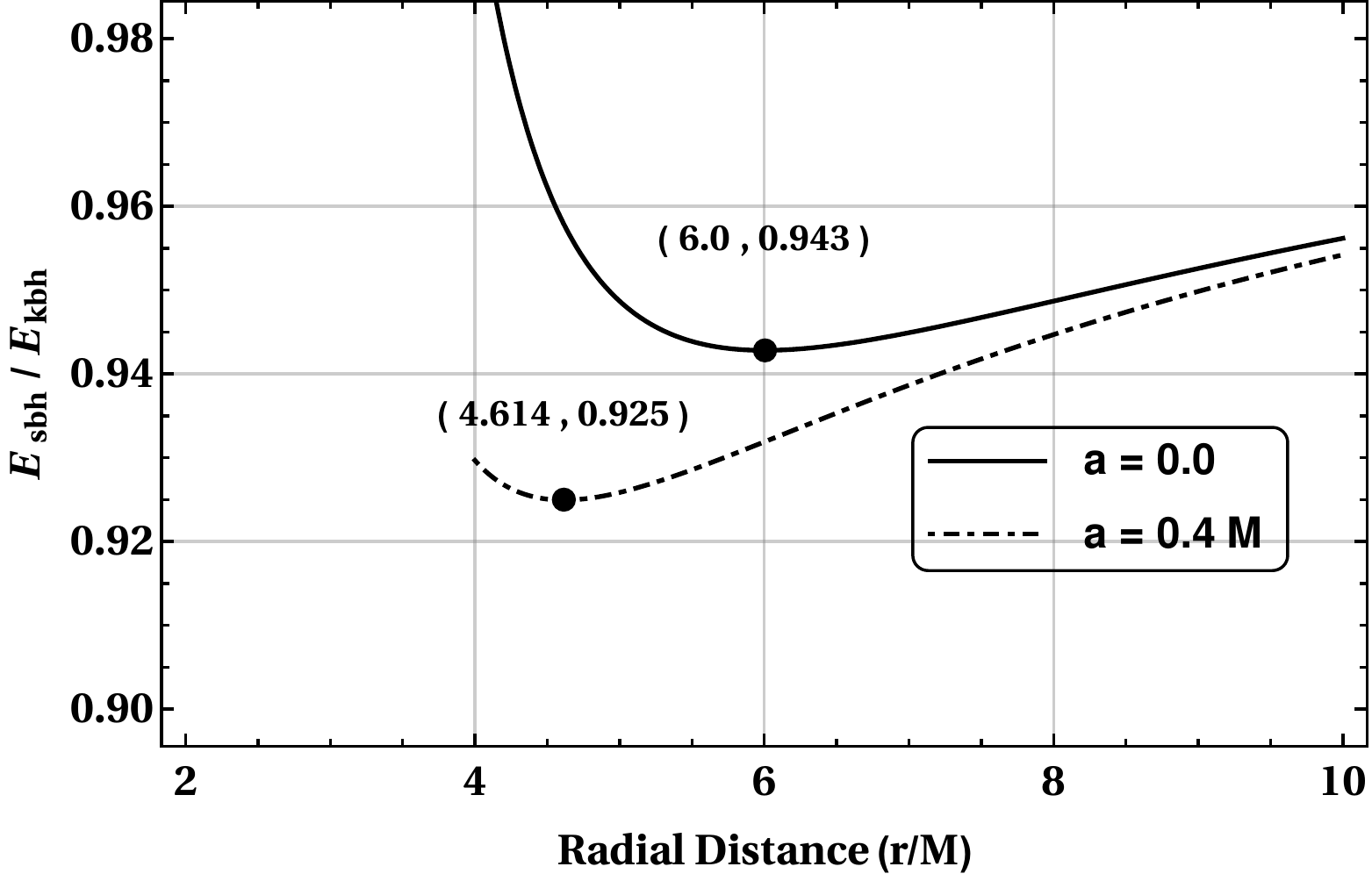}
\includegraphics[scale=.5]{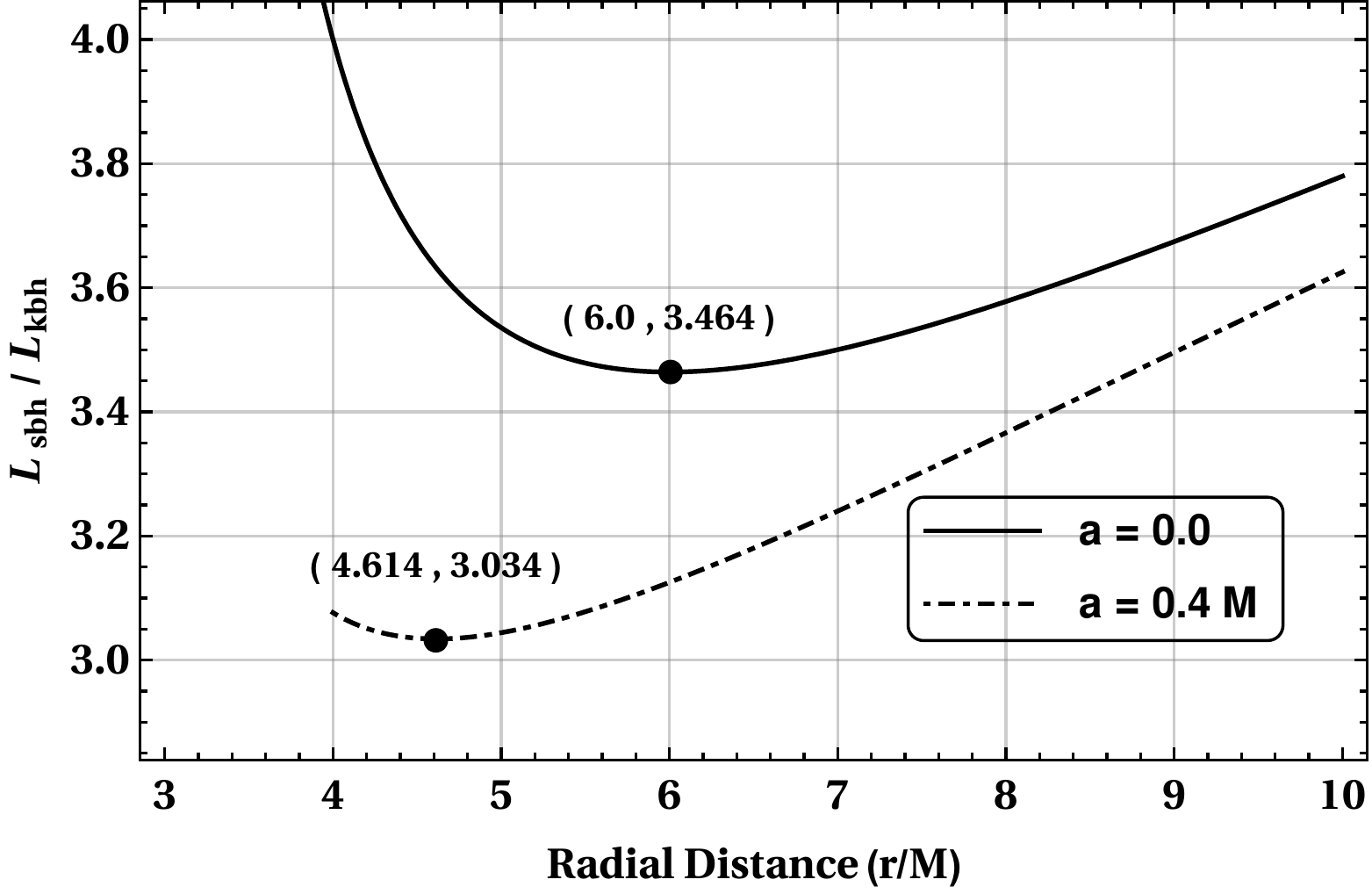}
\caption{Energy and angular momentum of a timelike geodesic is shown in a rotating as well as in a non-rotating gravitational field. The ISCO always appears at a point where energy and angular momentum both simultaneously become minimum.}
\label{SBH_GD}
\end{figure}

In this section, we shall discuss the stability of the non-equatorial circular orbits for a spinning particle. Before investigating the spinning particle, we revisit the stability properties of the geodesic trajectories around a Schwarzschild and Kerr black hole. In this case, the energy and angular momentum are easily derivable from the radial potential:
 \begin{equation}
 E_{\rm sbh}^2 = \dfrac{1}{r}\dfrac{(r-2 M)^2}{ (r-3 M)},  \qquad \text{and} \qquad L_{\rm sbh}^2 = \dfrac{Mr^2 }{(r-3 M)}.
\end{equation}  
Where $M$ is the mass of the black hole. Both energy and momentum reaches a simultaneous minima at $r=6M$, which is precisely the innermost stable circular orbit (ISCO) for a timelike particle \cite{Tsupko:2016bpn,Chaverri-Miranda:2017gxq}. Beyond this limit, no stable circular orbit is possible in a Schwarzschild spacetime. A similar situation appears around a Kerr black hole with energy $E_{\rm kbh}$ and momentum $L_{\rm kbh}$. But in that case, ISCO depends on the angular momentum of the black hole. For example, at $a=0.4M$, the ISCO for a co-rotating geodesic appears at $r=4.614 M$. A schematic diagram to demonstrate the ISCO is given in \ref{SBH_GD}. 

\begin{figure}[htp]
 \centering
 \includegraphics[scale=.48]{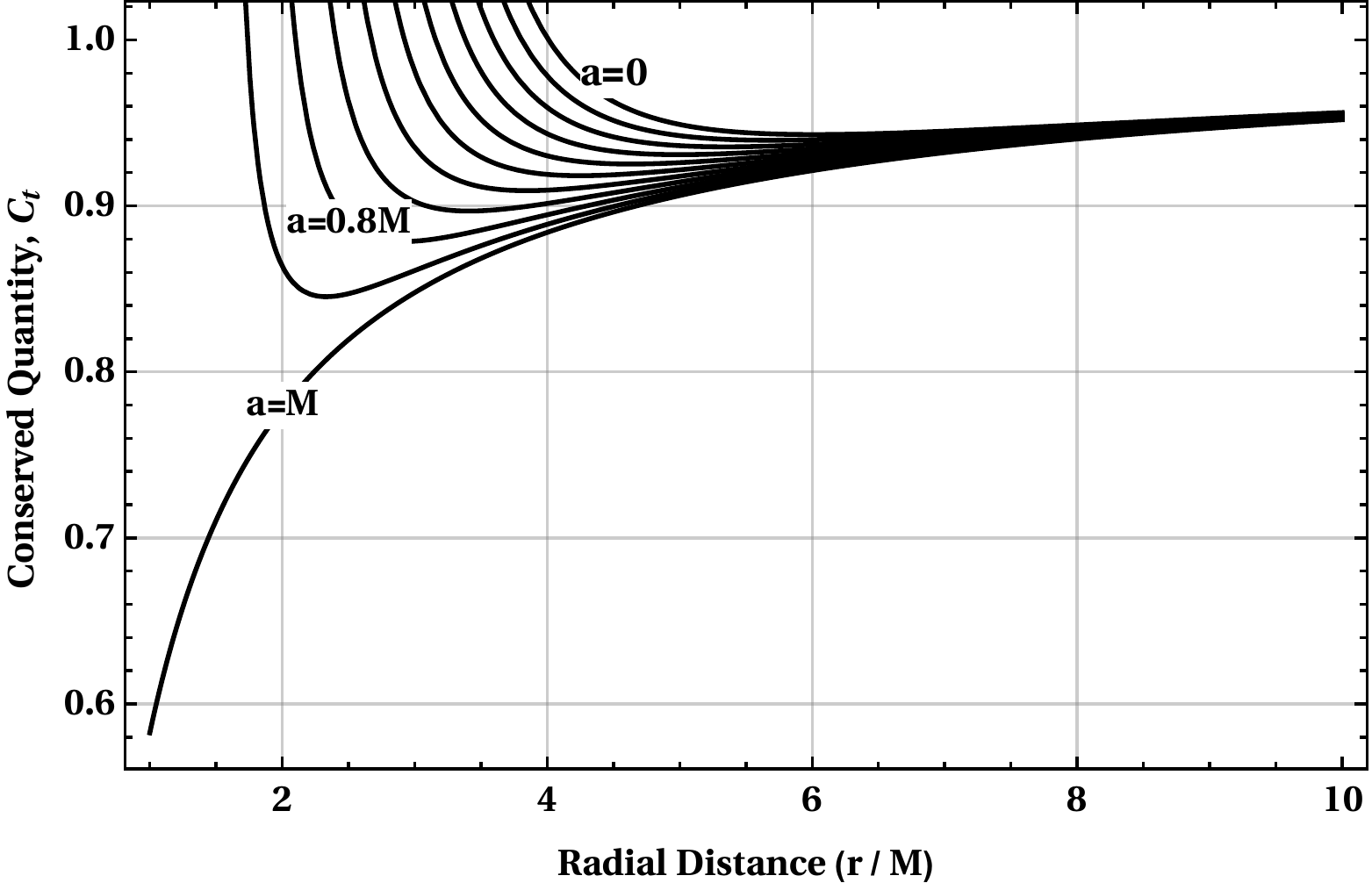}
  \includegraphics[scale=.48]{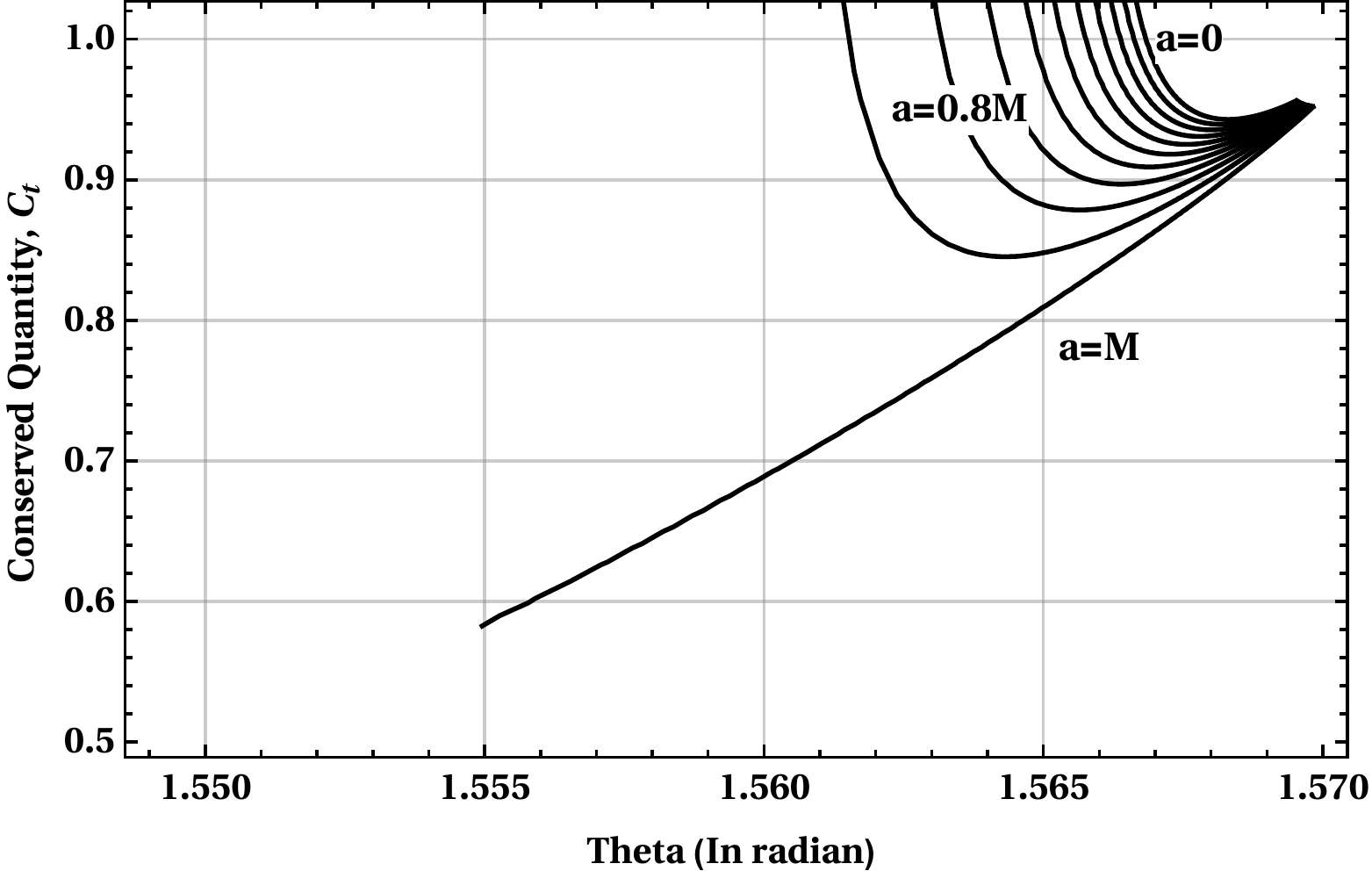}
  \includegraphics[scale=.48]{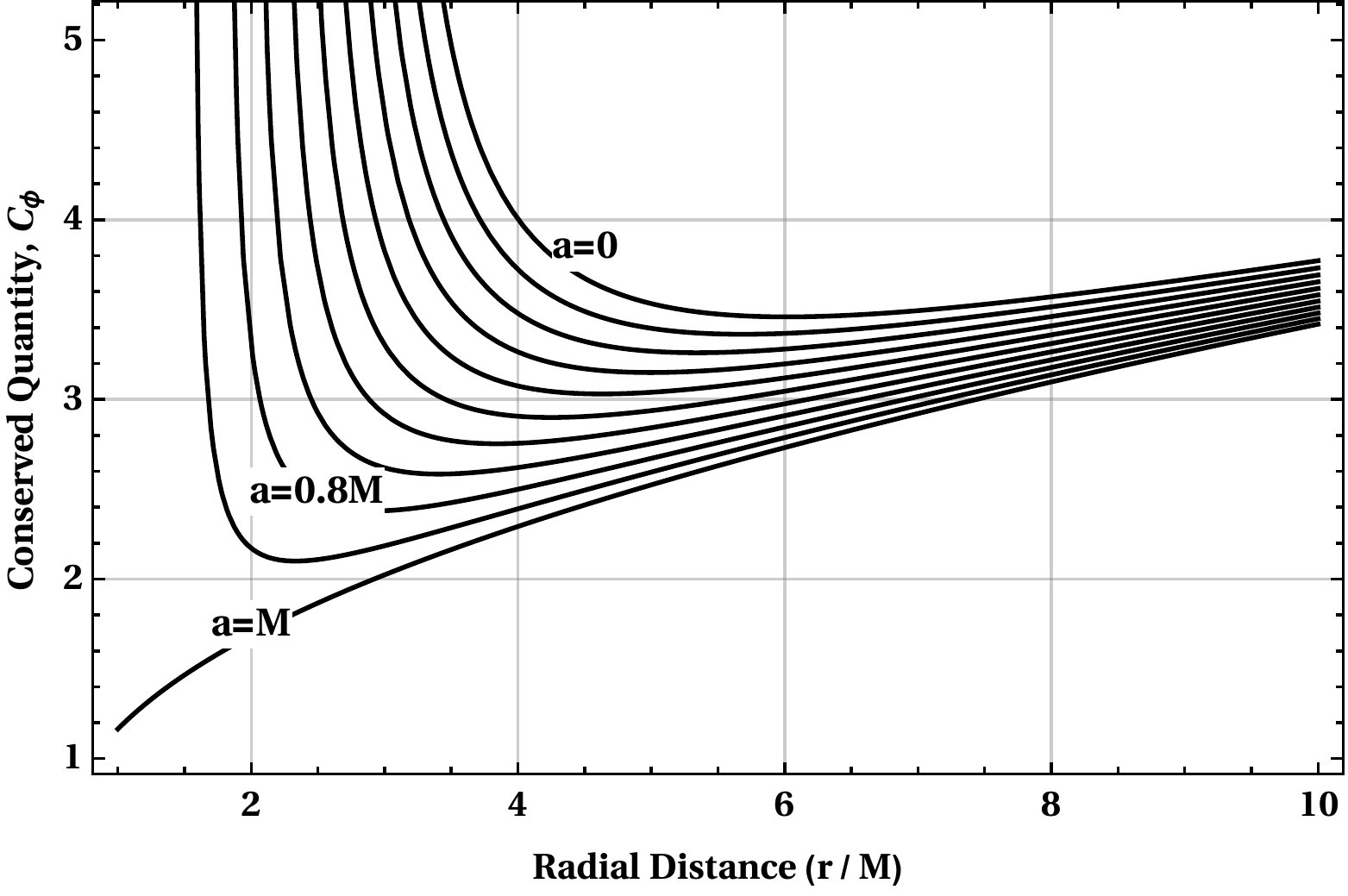}
  \includegraphics[scale=.48]{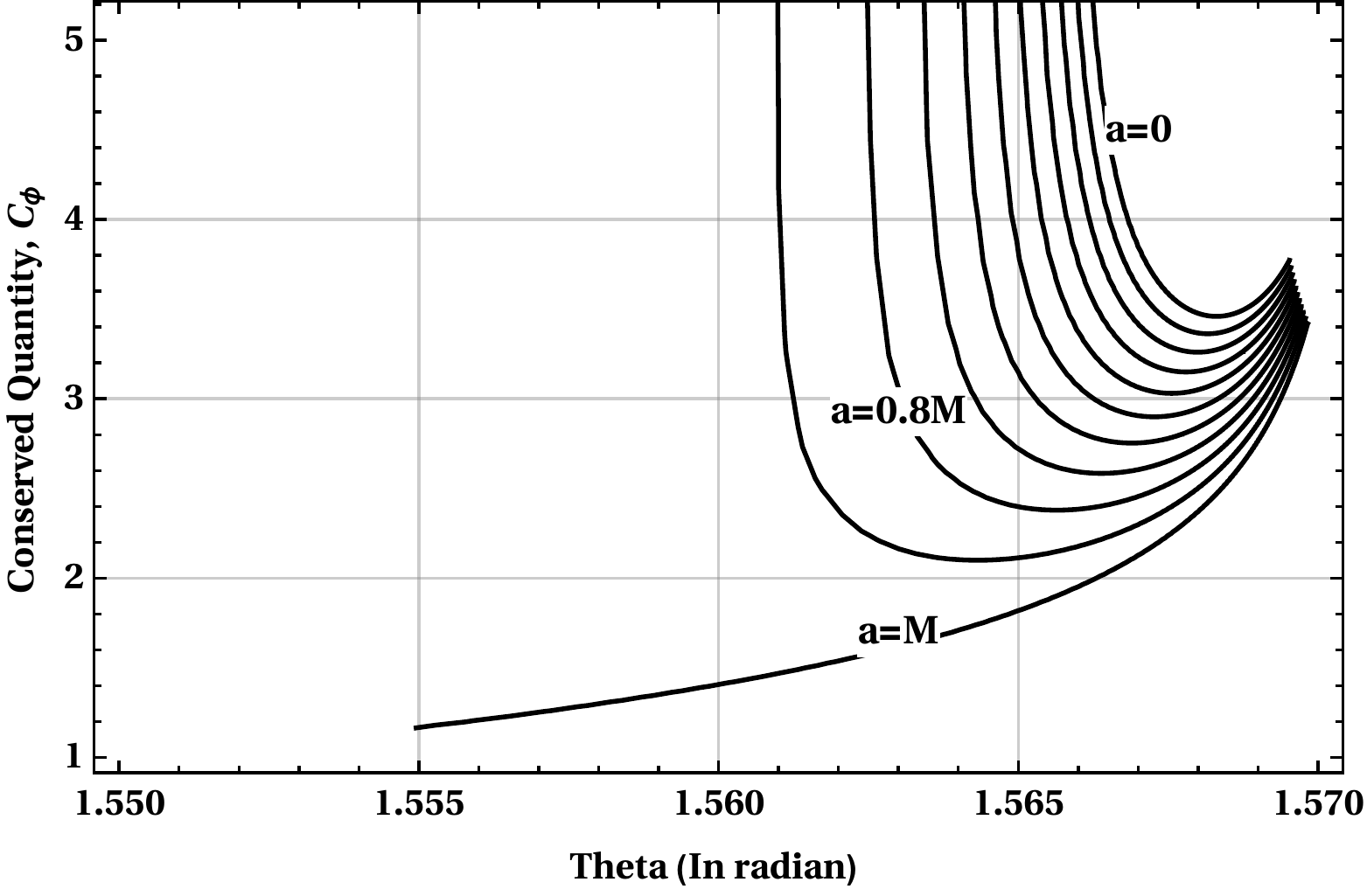}
  \caption{The conserved energy and total angular momentum is shown for co-rotating orbits with $S^{(1)}=-0.015M$ and $S^{(2)}=-0.01M$ in the \T~\SSC. The ISCO is shown as the minimum of energy and angular momentum, which is placed in a non-equatorial plane. The variation of ISCO is shown for different angular momentum of the black hole. Black hole's angular momentum is zero for the upper branch and $a=M$ for the lower branch. In between $a$ is increasing from top to bottom. It is shown that for large values of \enquote*{a}, the ISCO exist in a more deviated non-equatorial plane while for $a \rightarrow 0$, the ISCO exist very close to the equatorial plane.}
 \label{fig:Energy_TSSC} 
 \end{figure}

\begin{figure}[htp]
\subfloat[Above figure shows $\mathcal{C}_t$ for a spinning particle with $S^{(1)}=-0.015M$ and $S^{(2)}=-0.01M$. Close to $a=M$ limit, the diagrams are distinctly different compare to other values of the angular momentum.\label{fig:Nearly_Kerr_01}]{%
  \includegraphics[height=4.9cm,width=.49\linewidth]{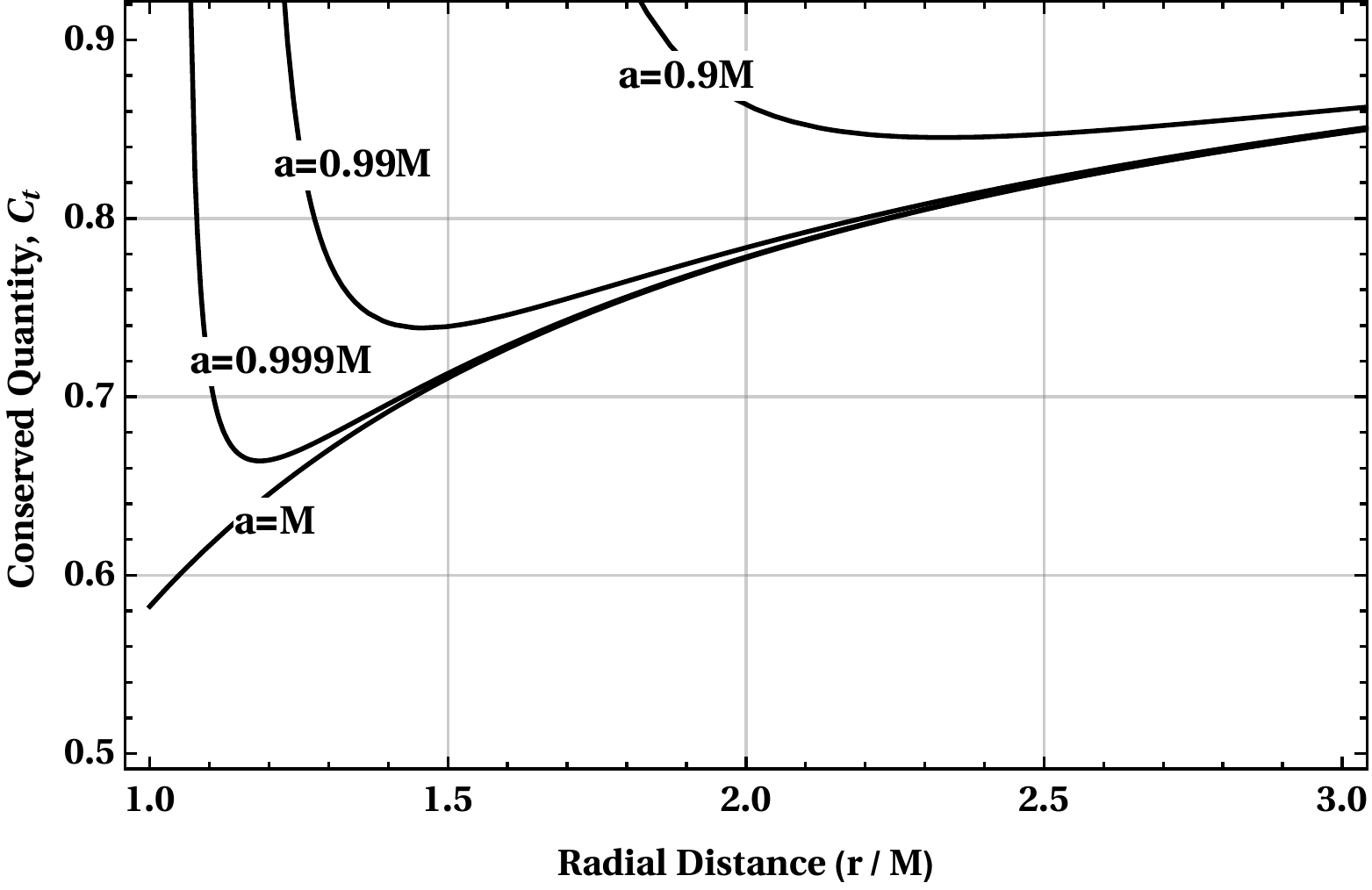}%
}\hfill
\subfloat[The conserved energy is shown for a geodesic in the Kerr black hole. For, $a=M$ the ISCO exists at $r=M$ for a co-rotating orbit. However, from the figure one can find that for $a=M$, the curve is sharply different from the rest.  \label{fig:Nearly_Kerr_02}]{%
  \includegraphics[height=4.9cm,width=.49\linewidth]{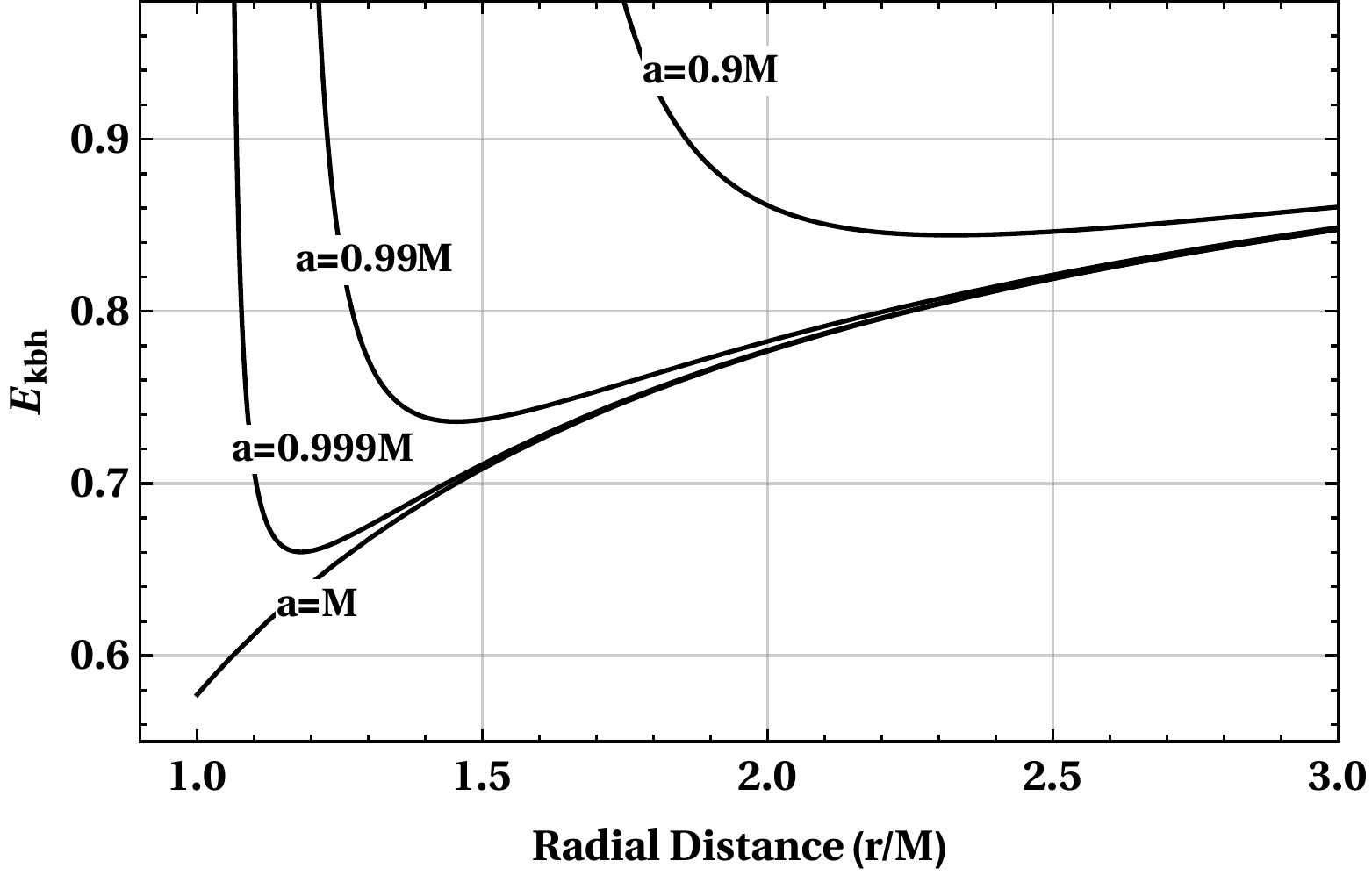}%
}
\caption{The conserved quantity is shown for a nearly extremal Kerr black hole for both spinning particle as well as geodesic.}
\label{fig:Nearly_Kerr}
\end{figure}

\begin{figure}[htp]
\centering
\includegraphics[scale=.48]{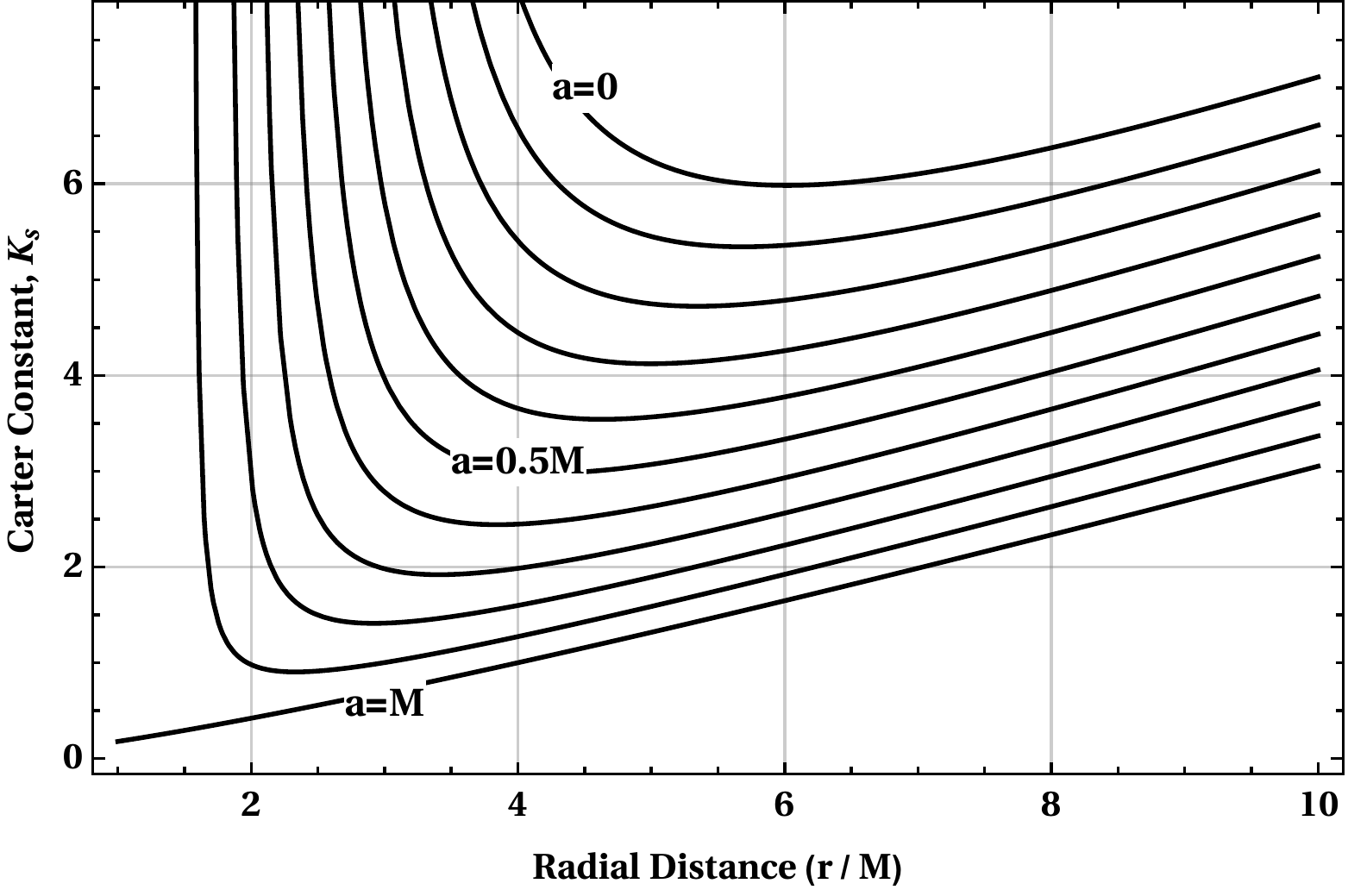}
\includegraphics[scale=.48]{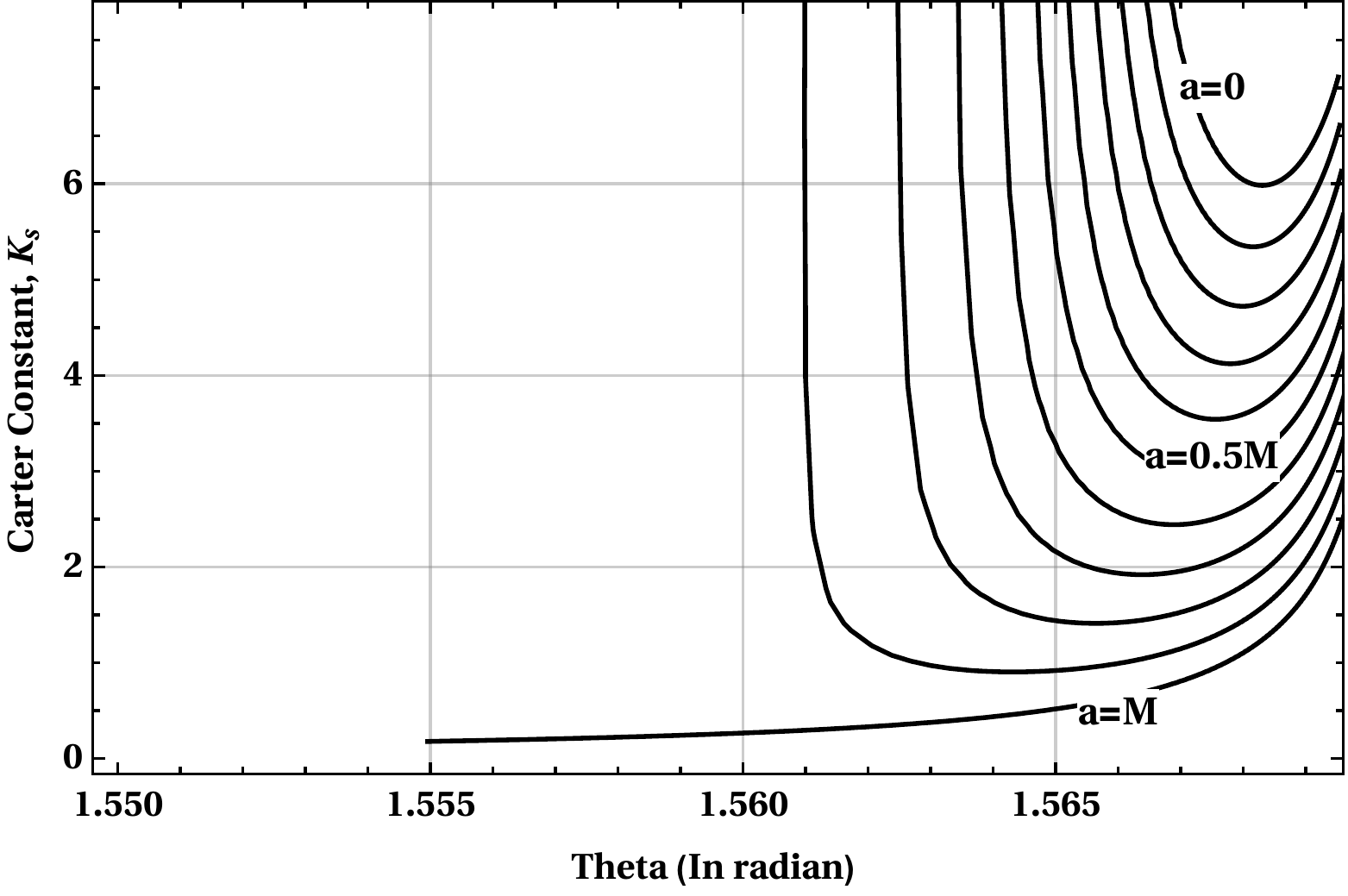}
\caption{The Carter constant is shown for $S^{(1)}=-0.015M$ and $S^{(2)}=-0.01M$ with the angular momentum of the geometry varies as shown in the figure. Similar to the other constants such as energy and angular momentum, Carter constant also reaches a minima in $r$ and $\theta$. It clearly suggests that the particle will eventually settle down in a non-equatorial orbit.}
\label{fig:Carter_constant}
\end{figure}

\begin{figure}[htp]
 \centering
 \includegraphics[scale=.48]{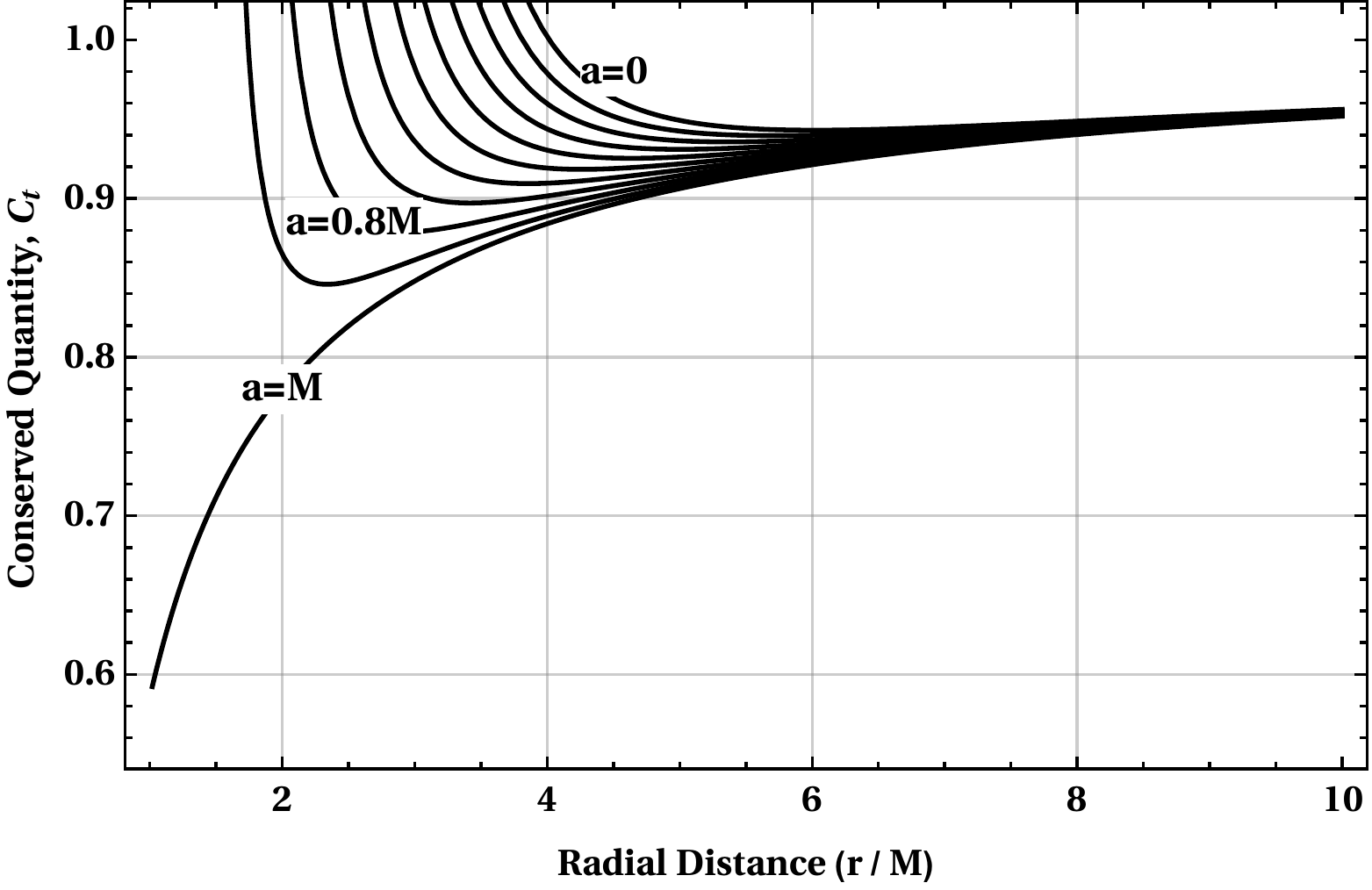}
  \includegraphics[scale=.48]{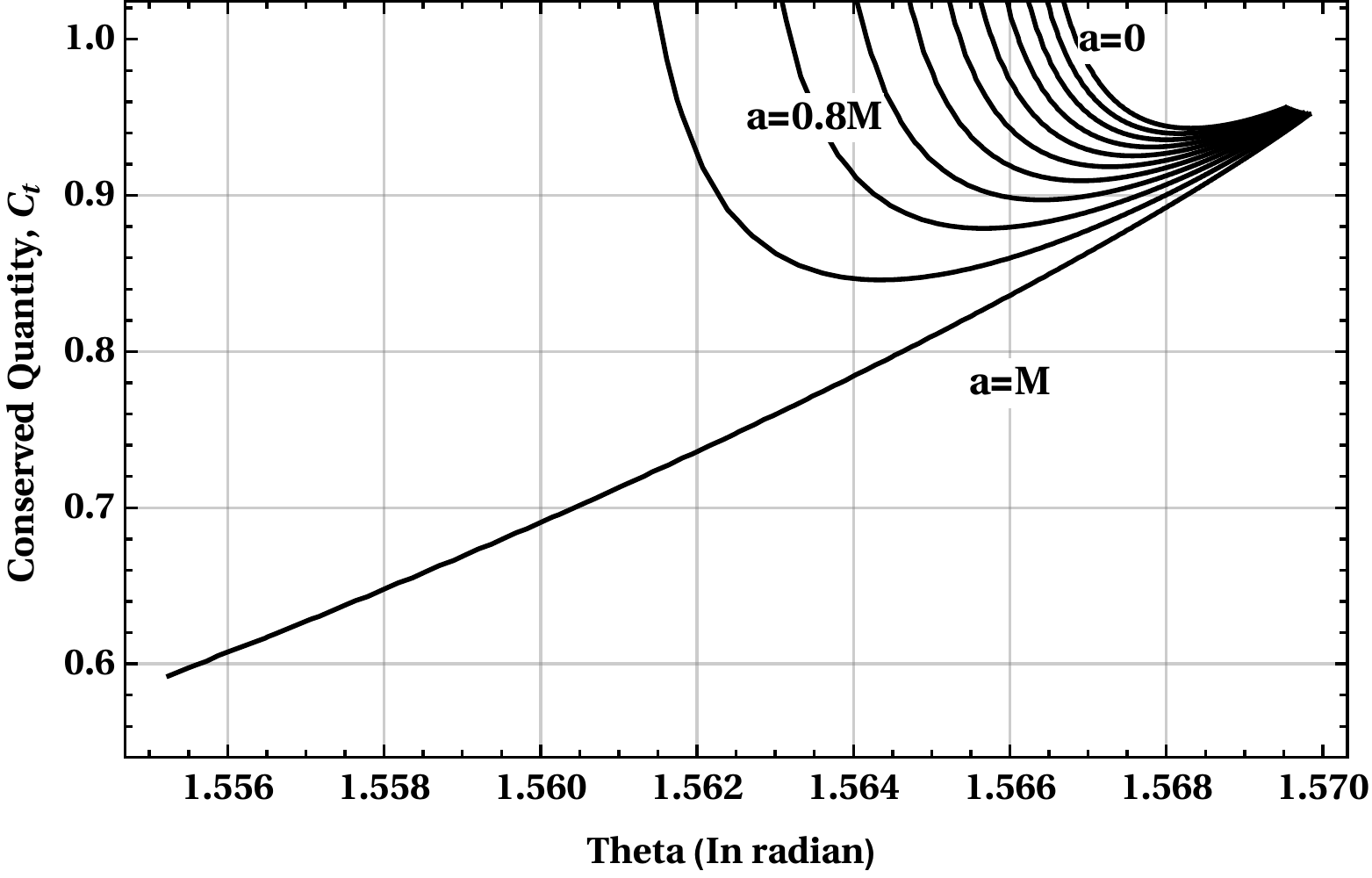}
  \includegraphics[scale=.48]{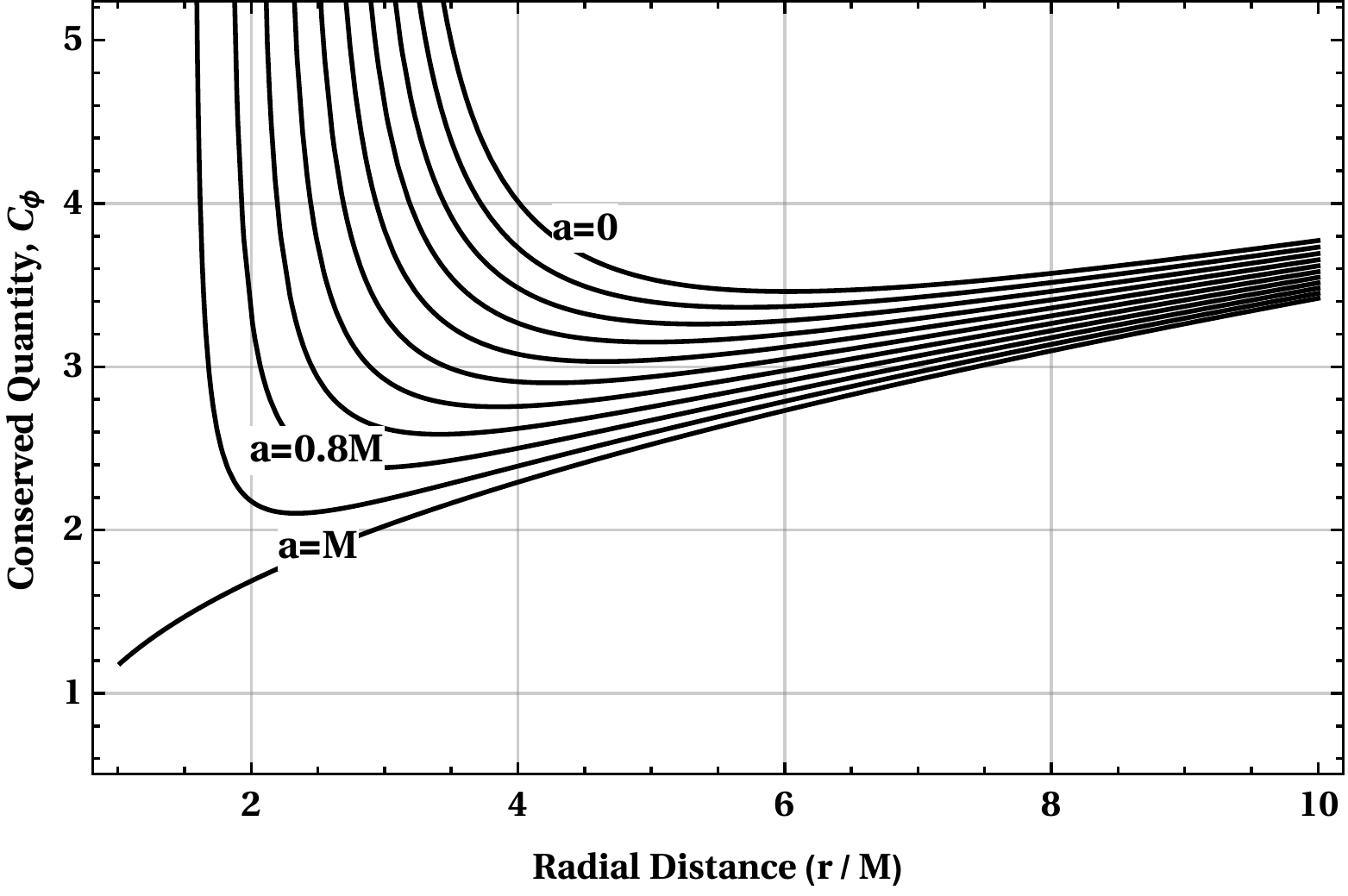}
  \includegraphics[scale=.48]{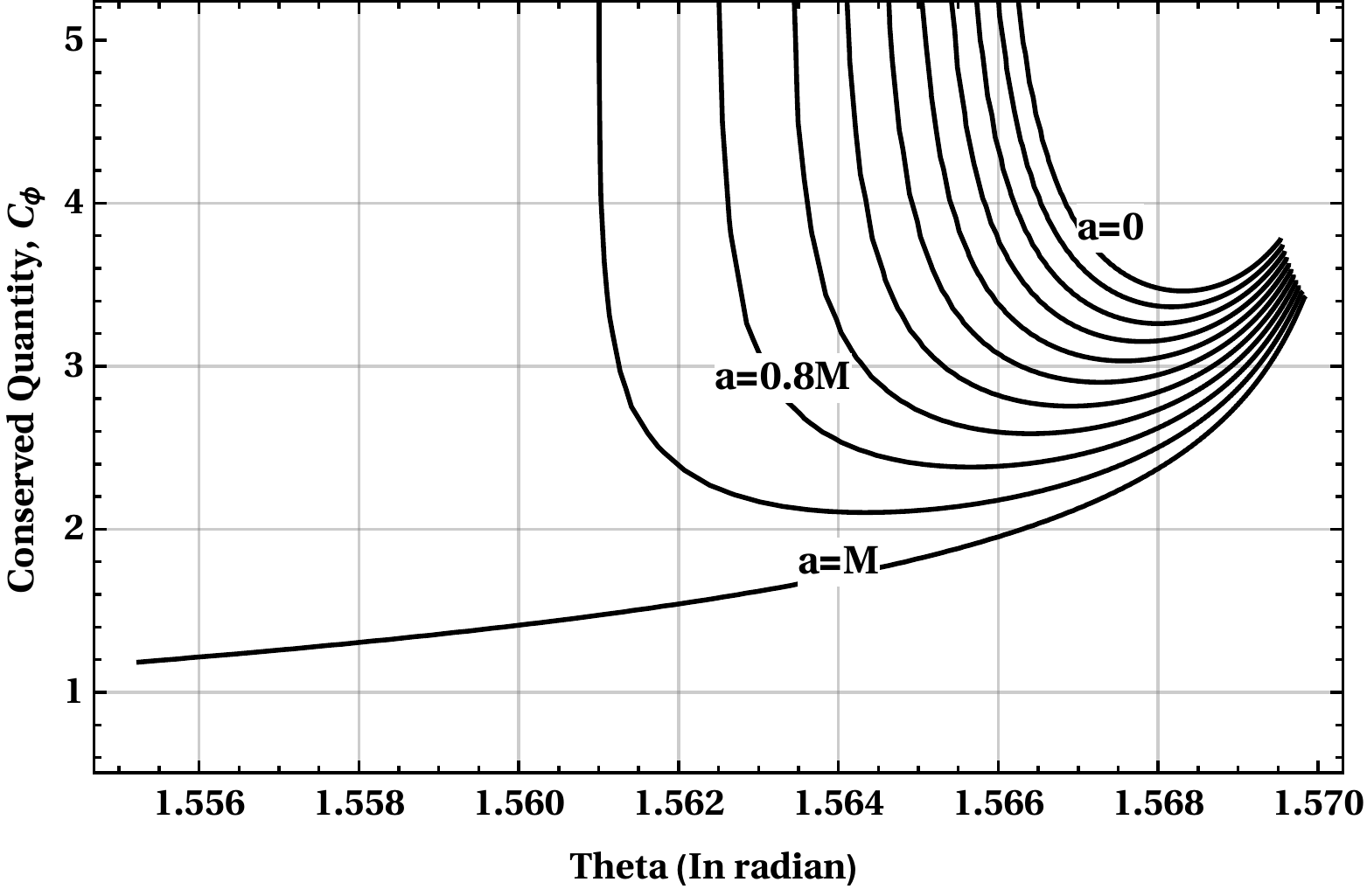}
  \caption{The conserved quantities are shown in the \NW~\SSC. Similar to the previous case of \T~\SSC, the minima appear in a non-equatorial plane.}
 \label{fig:Energy_NWSSC} 
 \end{figure}

\begin{figure}[htp]
\subfloat[ The radial coordinate for the \ISCO s are shown in the above figure for different angular momentum of the black hole while the spin parameters are fixed at $S^{(1)}=-0.015M$ and $S^{(2)}=-0.01M$.\label{fig:difference_1}]{%
  \includegraphics[height=4.9cm,width=.49\linewidth]{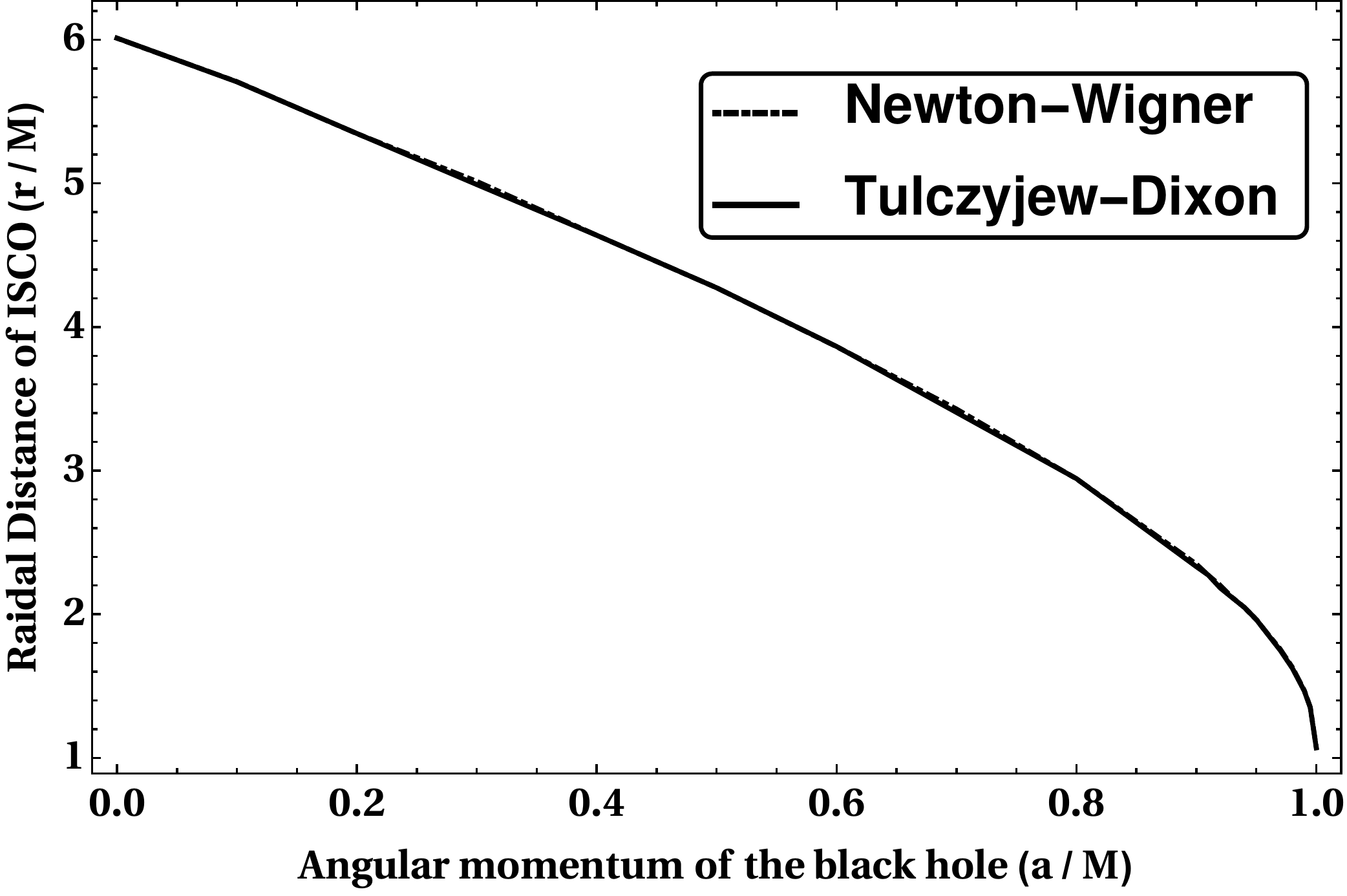}%
}\hfill
\subfloat[ The difference of \ISCO s for different \SSC s increases with the increase of spin parameters. In this case, we set $S^{(1)}=-0.05M$ and $S^{(2)}=-0.045M$. \label{fig:difference_1}]{%
  \includegraphics[height=4.9cm,width=.49\linewidth]{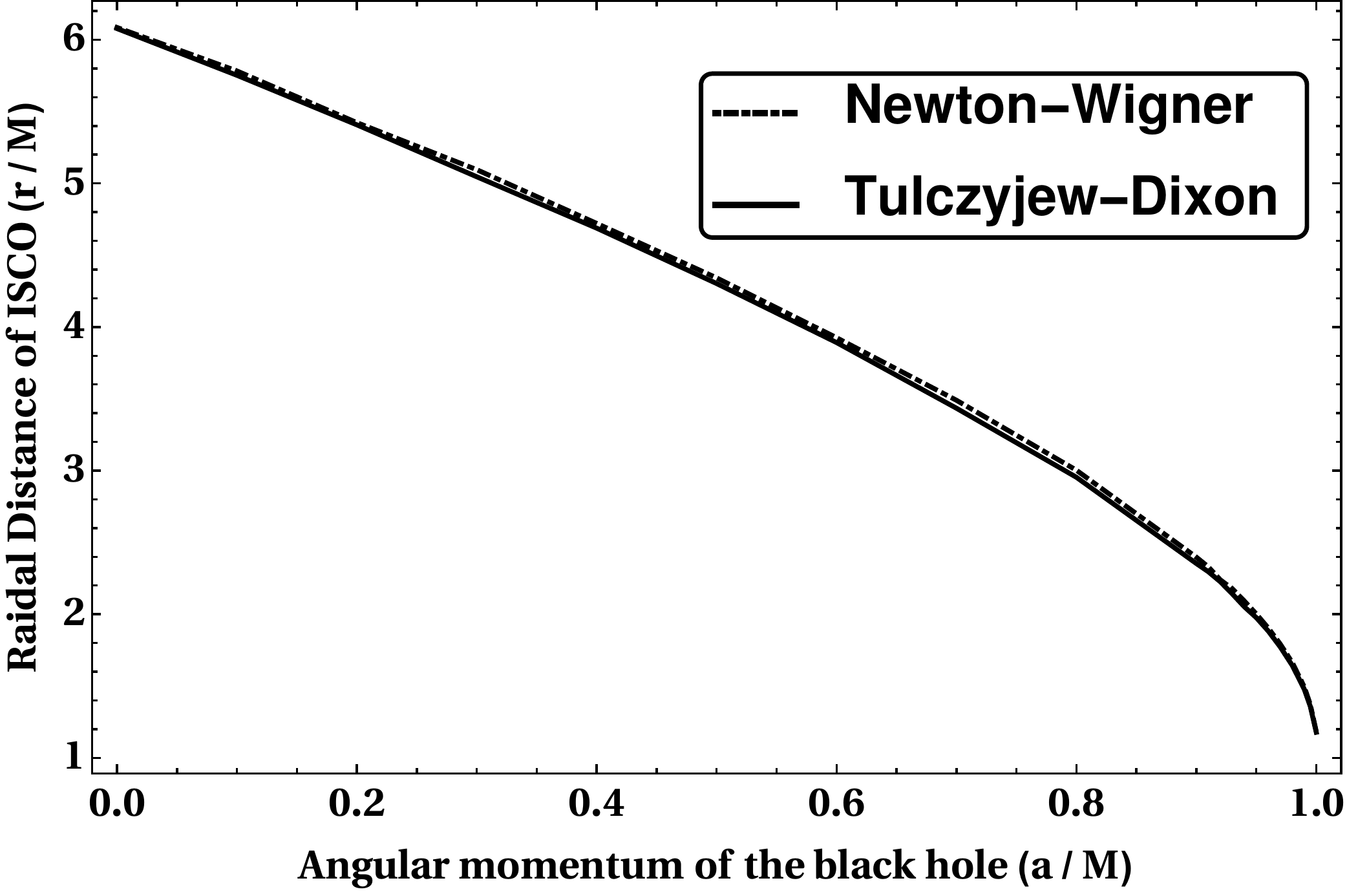}%
}
\caption{The \ISCO s are shown for different \SSC s. }
\label{fig:difference_02}
\end{figure}

\noindent
In case of a spinning particle, neither energy nor momentum is a conserved quantity, in stead, some spin dependent functions such as, $\mathcal{C}_t$, $\mathcal{C}_{\phi}$ and $K_s$ become constant of motion. Even then, one can further ask whether it is possible to have a simultaneous minima for these conserved quantities similar to a non-spinning particle, i.e, can the existence of ISCO be extended for a spinning particle as well. If possible, the conserved quantities  has to the satisfy the following equations at the ISCO located at $(r_c,\theta_c)$
\begin{equation}
\left(\dfrac{\partial \mathcal{C}_t}{\partial r}\right)_{r=r_c}=0, \qquad \left(\dfrac{\partial \mathcal{C}_{\phi}}{\partial r}\right)_{r=r_c}=0, \quad \text{and} \quad \left(\dfrac{\partial K_s}{\partial r}\right)_{r=r_c}=0,
\label{eq_minima_r}
\end{equation} 
\begin{equation}
\left(\dfrac{\partial \mathcal{C}_t}{\partial \theta}\right)_{\theta=\theta_c}=0, \qquad \left(\dfrac{\partial \mathcal{C}_{\phi}}{\partial \theta}\right)_{\theta=\theta_c}=0, \quad \text{and} \quad \left(\dfrac{\partial K_s}{\partial \theta}\right)_{\theta=\theta_c}=0.
\label{eq_minima_theta}
\end{equation} 
or can be stated the other way round, \textit{i.e}, if we could establish that the above sets of equations are satisfied at some $(r_c,\theta_c)$ then it has to be the ISCO for a spinning particle in the Kerr spacetime. But unfortunately, to find any analytical solutions of the above equations is a formidable task and one has to rely on some numerical techniques which we shall carry out in this section. To complete the task, we first introduce the velocity components $U^{(0)}$ and $U^{(3)}$ as
\begin{equation}
U^{(0)}=\dfrac{1}{\sqrt{1-\overline{\Omega}^2}}, \qquad \text{and} \qquad U^{(3)}=\dfrac{\overline{\Omega}}{\sqrt{1-\overline{\Omega}^2}}.
\end{equation}
 Substituting the values of $\overline{\Omega}$, $r$ and $\theta$, we plot the variation of $\mathcal{C}_t$, $\mathcal{C}_{\phi}$ and $K_s$, as shown in \ref{fig:Energy_TSSC} and \ref{fig:Carter_constant}. It can be noted that all of them has a simultaneous minima in a non-equatorial orbit. This corresponds to ISCO for the spinning particles and unlike geodesics, it appear in a off-equatorial plane. The ISCO for \NW~\SSC~is shown in \ref{fig:Energy_NWSSC} and the  comparison between two \SSC s~is depicted in \ref{fig:difference_02}. It is easy to notice that the difference for different \SSC s is small in the linear spin approximation. 
In addition, one may also interpret from \ref{fig:Energy_TSSC} that the behavior of both these conserved quantities are distinctly different for $a=M$ compare to other values of a. To clarify this point, we have shown $\mathcal{C}_t$ for different values of a in the range $0.9M$ to $M$, as given in \ref{fig:Nearly_Kerr_01}. Clearly for $a=M$, the familiar structure to find the ISCO from a graphical representation is absent and the minimum value of $\mathcal{C}_t$ is taken upto which the motion is described. This is a consequence of the spacetime geometry and can be shown to exist even for a geodesic trajectory, as it is depicted in \ref{fig:Nearly_Kerr_02}. Due to the  existence of the horizon at $r=M$ for an extremal black hole, the radial coordinate is not extensible beyond $r=M$ and that is exactly where the ISCO is located. This would essentially change the behavior of the plots near $a=M$ limit. Similar outcome can be anticipated even for a spinning particle.
\vspace{2cm}

\noindent
For a more rigorous proof, it is instructive to simultaneously solve \ref{eq_minima_r} and \ref{eq_minima_theta} and conclude that there indeed exists a common solution $(r_c,\theta_c)$ to these equations. To complete this task, we have employed the technique involving Lagrangian multiplier which is useful to compute any extrema of the function $f(x,y)$ while $x$ and $y$ further satisfy $g(x,y)=0$. Let us now define a new function $\mathcal{L}(x,y)$ such as
\begin{equation}
\mathcal{L}(x,y)=f(x,y)-\lambda g(x,y)
\end{equation}
with $\lambda$ is given as a constant which we indented to evaluate. To find the respective minima in both $x$ and $y$, the following expressions has to be satisfied
\begin{equation}
\dfrac{\partial \mathcal{L}(x,y)}{\partial x}=\dfrac{\partial f(x,y)}{\partial x}-\lambda \dfrac{\partial g(x,y)}{\partial x}=0, \quad \text{and} \quad \dfrac{\partial \mathcal{L}(x,y)}{\partial y}=\dfrac{\partial f(x,y)}{\partial y}-\lambda \dfrac{\partial g(x,y)}{\partial y}=0.
\end{equation}
Solving for $\lambda$, we may conclude the validity of the following expression
\begin{equation}
\dfrac{\partial f(x,y)}{\partial x}\dfrac{\partial g(x,y)}{\partial y}-\dfrac{\partial f(x,y)}{\partial y}\dfrac{\partial g(x,y)}{\partial x}=0,
\end{equation}
Therefore, it is easy to predict that whenever the above equation along with $g(x,y)=0$ is satisfied, there exists a minima to the function $f(x,y)$ at $(x_c,y_c)$. We use this technique in the present context with $f(x,y)$ corresponds to the conserved quantities and $g(x,y)$ is the constraint arrives from the orbit equation. These results are reproduced in \ref{Table_01} with respective mean and deviation for different momentum of the black hole. It can be easily noticed that there exists a common solution to \ref{eq_minima_r} and \ref{eq_minima_theta} at $r_c$ and $\theta_c$ within a maximum error of $\mathcal{O}(10^{-5})$. The mean $\overline{r_c}$ used in \ref{Table_01} is given as $\overline{r}_c=\sum_{i=1}^{3} \dfrac{r^{i}_{c}}{3}$ and the deviation is defined as $<r_c> = \sqrt{\sum_{i=1}^{3} \dfrac{(\overline{r}_c-r^{i}_{c})^2}{3}}$, where $r^{i}_{c}$ corresponds to different solutions obtained for various equations relating $E$, $J_z$ and $K_s$. Similarly, one can estimate $\overline{\theta}_c$ and $<\theta_c>$ and show that the solutions coincide within an error bound of $\mathcal{O}(10^{-8})$.

\begin{table}[htp]
 \begin{adjustwidth}{-1cm}{}
\begin{center}
\caption{The solutions to \ref{eq_minima_r} and \ref{eq_minima_theta} are computed numerically for various momentum parameters of the black hole and the spin vector follows $S^{(1)}=-0.015M$ and $S^{(2)}=-0.01M$.} 
\label{Table_01}
\vskip 2mm
\begin{tabular}{*{6}{|c}|}
\hline
\multicolumn{1}{|c|}{}  & \multicolumn{5}{c|}{Angular momentum of the black hole}  \\ 
\hline
\multicolumn{1}{|c|}{Expressions}  & $a=0$ & $a=0.3M$ &  $a=0.5M$ & $a=0.7M$ & $a=0.9M$  \\ 
\hline
\multicolumn{1}{|c|}{$d\mathcal{C}_t/dr=0$}  & $r_c=6.01632 M$ & $r_c=4.99410 M$ & $r_c=4.24766 M$ & $r_c=3.40649 M$ & $r_c=2.33143M$  \\
                                 & $\theta(r_c)=1.56831$ & $\theta(r_c)=1.56780$ & $\theta(r_c)=1.56727$ & $\theta(r_c)=1.56639$ & $\theta(r_c)=1.56432$\\
\hline 
\multicolumn{1}{|c|}{$d\mathcal{C}_{\phi}/dr=0$} & $r_c=6.01637M$  & $r_c=4.99416M$ & $r_c=4.24772M$ & $r_c=3.40655M$ & $r_c=2.33149M$  \\
								  & $\theta(r_c)=1.56831$ & $\theta(r_c)=1.56780$ & $\theta(r_c)=1.56727$ & $\theta(r_c)=1.56639$ & $\theta(r_c)=1.56432$ \\
\hline 
\multicolumn{1}{|c|}{$dK_s/dr=0$} & $r_c=6.01631M$ & $r_c=4.99408M$ & $r_c=4.24763M$ & $r_c=3.40645M$ & $r_c=2.33137M$  \\
								& $\theta(r_c)=1.56831$ & $\theta(r_c)=1.56780$ & $\theta(r_c)=1.56727$ & $\theta(r_c)=1.56639$ & $\theta(r_c)=1.56432$ \\
\hline 
\multicolumn{1}{|c|}{$d\mathcal{C}_t/d\theta=0$} & $\theta_c=1.56831$ & $\theta_c=1.56780$ & $\theta_c=1.56727$  & $\theta_c=1.56639$  & $\theta_c=1.56430$ \\
									& $r(\theta_c)=6.01632M$ & $r(\theta_c)=4.99410M$ & $r(\theta_c)=4.24766M$ & $r(\theta_c)=3.40649M$ & $r(\theta_c)=2.33142M$ \\
\hline
\multicolumn{1}{|c|}{$d\mathcal{C}_{\phi}/d\theta=0$} & $\theta_c=1.56831$ & $\theta_c=1.56780$ & $\theta_c=1.56727$ & $\theta_c=1.56639$ & $\theta_c=1.56430$  \\
									& $r(\theta_c)=6.01637M$ & $r(\theta_c)=4.99416M$ & $r(\theta_c)=4.24772M$ & $r(\theta_c)=3.40655M$ & $r(\theta_c)=2.33149M$ \\ 
\hline
\multicolumn{1}{|c|}{$dK_s/d\theta=0$} & $\theta_c=1.56831$ & $\theta_c=1.56780$ & $\theta_c=1.56727$ & $\theta_c=1.56639$ & $\theta_c=1.56430$  \\ 
									& $r(\theta_c)=6.01632 M$ & $r(\theta_c)=4.99408 M$ & $r(\theta_c)=4.24763 M$ & $r(\theta_c)=3.40644 M$ & $r(\theta_c)=2.33137 M$ \\
\hline
\multicolumn{1}{|c|}{\thead{Mean}} & $\overline{r_c}=6.01633 M$ & $\overline{r_c}=4.99411 M$  & $\overline{r_c}=4.24767 M$  & $\overline{r_c}=3.4065 M$  & $\overline{r_c}=2.33143 M$  \\ 
						    & $\overline{\theta_c}=1.56831$ & $\overline{\theta_c}=1.56780$ & $\overline{\theta_c}=1.56727$ & $\overline{\theta_c}=1.56639$ & $\overline{\theta_c}=1.56430$ \\
\hline
\multicolumn{1}{|c|}{\thead{Deviation \\ (in $M^{-1}$ units)}} & $<r_c>=2.6 \times 10^{-5} $ & $<r_c>=3.4 \times 10^{-5} $  & $<r_c>=3.8 \times 10^{-5} $ & $<r_c>=4.4 \times 10^{-5} $ & $<r_c>=4.7 \times 10^{-5} $ \\ 
						    & $<\theta_c>=1.5 \times 10^{-8}$ & $<\theta_c>=2.5 \times 10^{-8}$ & $<\theta_c>=4 \times 10^{-8}$ & $<\theta_c>=7.2 \times 10^{-8}$ & $<\theta_c>=1.7 \times 10^{-7}$\\
\hline						    
						    
\end{tabular}
\end{center}
 \end{adjustwidth}
\end{table}
\noindent
For a consistency check, it should be reminded that for very small spin values, the ISCO should approximately match with the usual geodesic trajectories on the equatorial plane of a Kerr black hole. For example, with the spin parameter $S=(0,-1.5 \times 10^{-4}M,-1.0 \times 10^{-4}M , 0 )$ and a black hole with spin $a=0.5M$, if we conduct the similar prescription as stated above we found $\overline{r_c}=4.23315 M$ and $\overline{\theta_c}=1.57076$ within the error bounds $<r_c>=3.8 \times 10^{-9}M$ and $<\theta_c>=3.7 \times 10^{-14}$ respectively. However, the estimated radius of the innermost stable circular orbit for a geodesic exists at $r_{\rm isco}=4.233M$ on the equatorial plane of the black hole with rotation parameter $a=0.5M$. This demonstrates that for a spinning particle with $S \approx \mathcal{O}(10^{-4}M)$, the innermost stable circular orbits exist on a plane within a difference $\mathcal{O}(10^{-5})$ from the equatorial plane while the radial distance located within a difference of $\mathcal{O}(10^{-4})$ from the usual geodesic orbit. By further decrease in the spin parameter, $r_c$ and $\theta_c$ would match even more closely with the given ISCO for any geodesic trajectory.


\section{Discussion}\label{sec:Discussion}
The circular motion of spinning particles are discussed on the $\theta=\text{constant}$ plane in a  Kerr background. We numerically solve the \MP~equations and explicitly shown the existence of such orbits in a rotating geometry for different \SSC s. The deviation from the equatorial plane is proportional to the radial spin component ($S^r$) of the particle as well as the angular momentum of the black hole. But the direction of the deviation is related to the sign of $S^{(1)}\overline{\Omega}$. More precisely, as shown in \ref{fig:Non_Eq}, for $S^{(1)}<0$ the counter-rotating orbits deviate in the direction of $\theta=\pi$, while an opposite phenomena appear for co-rotating orbits.
The study of different \SSC s are carried out in the linear spin framework. The nature of the plots remain similar as shown in \ref{fig:Non_Eq} and \ref{fig:Non_Eq_NW} , and the difference is very small for different \SSC s. 
We provided a better understanding of the stabilities of these circular orbits in terms of conserved quantities such as energy, angular momentum and Carter constant. Similar to the geodesics, there exist a point ($r_c,\theta_c$) where all the conserved quantities reaches their respective minima simultaneously and this corresponds to the ISCO. This suggests that spinning particles not only can move in the non-equatorial circular orbits, they may even settle down in such planes. Due to such interesting consequences it may be possible to detect those orbits in real astrophysical domain. Motion of extended objects around the Supermassive black hole Sgr A* in our Galaxy \cite{Schodel:2002vg,Hees:2017aal} or the extreme mass ratio binaries may serve as promising candidates to witness any imprints of off-equatorial stable circular orbits for a spinning particle. Furthermore, as the complete analysis is carried out within the linear spin limit, the results produced in the article could be easily employed for any astrophysical event as far as the spin vector satisfies $S \ll \mathcal{O}(M)$. However, it would be intriguing from theoretical perspective to search for any possibilities of such orbits whenever the linear constraint is relaxed and $\mathcal{O}(S^2)$ are also considered.

 \section{Acknowledgement}
The authors extend their gratitude to the Inter-University Centre
for Astronomy and Astrophysics (IUCAA), Pune for using their library and computation facilities during a visit under their associateship programme. They also wish to thank the anonymous referee for some constructive suggestions to improve the manuscript.

 \bibliographystyle{utphys1}
\bibliography{References}

 \end{document}